\documentclass[longauth]{aa}
\pdfoutput=1

\usepackage{txfonts}
\usepackage{xcolor}
\usepackage{graphicx}
\usepackage{array}
\usepackage{amsmath}
\usepackage{float}
\usepackage{hyperref}
\usepackage{ulem}
\usepackage{makecell}
\usepackage{upgreek}

\makeatletter
\renewcommand*\aa@pageof{, page \thepage{} of \pageref*{LastPage}}
\bibpunct{(}{)}{;}{a}{}{,} 
\newcommand{\vc}[1]{#1}
\newcommand{\new}[1]{#1}

\hypersetup{
    colorlinks,
    linkcolor={red!50!black},
    citecolor={blue!50!black},
    urlcolor={blue!80!black}
}

\begin{document}

\title{MINDS: JWST/NIRCam imaging of the protoplanetary disk PDS~70}
\subtitle{A spiral accretion stream and a potential third protoplanet}
\titlerunning{JWST/NIRCam imaging of the protoplanetary disk PDS~70}

\authorrunning{V.~Christiaens et al.}
\author{V.~Christiaens\inst{\ref{KULeuven},\ref{STARLiege}}, 
M.~Samland\inst{\ref{MPIA}}, 
Th.~Henning\inst{\ref{MPIA}}, 
B.~Portilla-Revelo\inst{\ref{Kapteyn}}, 
G.~Perotti\inst{\ref{MPIA}}, 
E.~Matthews\inst{\ref{MPIA}}, 
O.~Absil\inst{\ref{STARLiege}}, 
L.~Decin\inst{\ref{KULeuven}}, 
I.~Kamp\inst{\ref{Kapteyn}}, 
A.~Boccaletti\inst{\ref{LESIA}}, 
B.~Tabone\inst{\ref{SaclayCNRSOrsay}},
G.-D.~Marleau\inst{\ref{Duisburg},\ref{Tuebingen},\ref{MPIA},\ref{Bern}},
E.~F.~van Dishoeck\inst{\ref{Leiden},\ref{MPE}}, M.~G\"udel\inst{\ref{AstroWien},\ref{ETHZ}}, P.-O.~Lagage\inst{\ref{SaclayCEAGif}}, 
D.~Barrado\inst{\ref{CABMadrid}}, A.~Caratti o Garatti\inst{\ref{INAFNapoli},\ref{IASDublin}}, 
A.~M.~Glauser\inst{\ref{ETHZ}}, G.~Olofsson\inst{\ref{AlbaNova}}, 
T.~P.~Ray\inst{\ref{IASDublin}}, S.~Scheithauer\inst{\ref{MPIA}}, B.~Vandenbussche\inst{\ref{KULeuven}}, L.~B.~F.~M.~Waters\inst{\ref{Radboud},\ref{SRON}},
A.~M.~Arabhavi\inst{\ref{Kapteyn}}, 
S.~L.~Grant\inst{\ref{MPE}}, H.~Jang\inst{\ref{Radboud}}, J.~Kanwar\inst{\ref{Kapteyn},\ref{IWFGraz},\ref{TUGraz}}, 
J.~Schreiber\inst{\ref{MPIA}}, K.~Schwarz\inst{\ref{MPIA}}, M.~Temmink\inst{\ref{Leiden}}, 
G.~\"Ostlin\inst{\ref{AlbaNova}}
}
    \institute{
    Institute of Astronomy, KU Leuven, Celestijnenlaan 200D, Leuven, Belgium \\
    \email{valentin.christiaens@kuleuven.be}
    \label{KULeuven}
    \and
    STAR Institute, Universit\'e de Li\`ege, All\'ee du Six Ao\^ut 19c, 4000 Li\`ege, Belgium
    \label{STARLiege}
    \and
    Max-Planck-Institut f\"{u}r Astronomie (MPIA), K\"{o}nigstuhl 17, 69117 Heidelberg, Germany
    \label{MPIA}
    \and
    Kapteyn Astronomical Institute, Rijksuniversiteit Groningen, Postbus 800, 9700AV Groningen, The Netherlands
    \label{Kapteyn}
    \and
    LESIA, Observatoire de Paris, Universit\'e PSL, CNRS, Sorbonne Universit\'e, Universit\'e de Paris, 5 place Jules Janssen, 92195 Meudon, France
    \label{LESIA}
    \and
    Universit\'e Paris-Saclay, CNRS, Institut d'Astrophysique Spatiale, 91405, Orsay, France
    \label{SaclayCNRSOrsay}
    \and
    Fakult\"at f\"ur Physik, Universit\"at Duisburg-Essen, Lotharstra\ss{}e 1, 47057 Duisburg, Germany
    \label{Duisburg}
    \and
    Institut f\"ur Astronomie und Astrophysik, Universit\"at T\"ubingen, Auf der Morgenstelle 10, 72076 T\"ubingen, Germany
    \label{Tuebingen}
    \and
    Physikalisches Institut, Universit\"{a}t Bern, Gesellschaftsstr.~6, 3012 Bern, Switzerland
    \label{Bern}
    \and
    Leiden Sterrewacht, Leiden University, 2300 RA Leiden, The Netherlands
    \label{Leiden}
    \and
    Max-Planck Institut f\"{u}r Extraterrestrische Physik (MPE), Giessenbachstr.~1, 85748, Garching, Germany
    \label{MPE}
    \and
    Institut f\"ur Astrophysik, Universit\"at Wien, T\"urkenschanzstr.~17, A-1180 Vienna, Austria
    \label{AstroWien}
    \and
    ETH Z\"urich, Institute for Particle Physics and Astrophysics, Wolfgang-Pauli-Str.~27, 8093 Z\"urich, Switzerland
    \label{ETHZ}
    \and
    Universit\'e Paris-Saclay, Universit\'e Paris Cit\'e, CEA, CNRS, AIM, F-91191 Gif-sur-Yvette, France
    \label{SaclayCEAGif}
    \and
    Centro de Astrobiolog\'ia (CAB), CSIC-INTA, ESAC Campus, Camino Bajo del Castillo s/n, 28692 Villanueva de la Ca\~nada, Madrid, Spain
    \label{CABMadrid}
    \and
    INAF – Osservatorio Astronomico di Capodimonte, Salita Moiariello 16, 80131 Napoli, Italy
    \label{INAFNapoli}
    \and
    Dublin Institute for Advanced Studies, 31 Fitzwilliam Place, D02 XF86 Dublin, Ireland
    \label{IASDublin}
    \and
    Institutionen f\"or Astronomi, Stockholms Universitet, AlbaNova Universitetscentrum, 10691 Stockholm, Sweden
    \label{AlbaNova}
    \and
    Department of Astrophysics/IMAPP, Radboud University, PO Box 9010, 6500 GL Nijmegen, The Netherlands
    \label{Radboud}
    \and
    SRON Netherlands Institute for Space Research, Niels Bohrweg 4, NL-2333 CA Leiden, The Netherlands
    \label{SRON}
    \and
    Institut f\"ur Weltraumforschung, Austrian Academy of Sciences, Schmiedlstr.~6, A-8042, Graz, Austria
    \label{IWFGraz}
    \and
    TU Graz, Fakult\"at für Mathematik, Physik und Geod\"asie, Petersgasse~16 8010 Graz, Austria
    \label{TUGraz}
    }
    %
    
    \date{Received 22 December 2023, Accepted 6 March 2024}
    
    \abstract
    {Two protoplanets have recently been discovered within the PDS~70 protoplanetary \new{disk}. 
    JWST/NIRCam offers a unique opportunity to characterize them and their birth environment at wavelengths that are difficult to access from the ground.}
    {We image the circumstellar environment of PDS~70 
    at 1.87\,$\upmu$m and 4.83\,$\upmu$m, assess the presence of Pa-$\alpha$ emission due to accretion onto the protoplanets, and probe any IR excess indicative of heated circumplanetary material.}
    {We obtained noncoronagraphic JWST/NIRCam images of PDS~70 within the MIRI mid-INfrared Disk Survey (MINDS) program. We leveraged the Vortex Image Processing (VIP) package for data reduction, and we developed dedicated routines for optimal stellar point spread function subtraction, unbiased imaging of the disk, and protoplanet flux measurement in this type of dataset. 
    A radiative transfer model of the disk was used to separate the 
    contributions from the disk and the protoplanets.}
    {We redetect both protoplanets
    and identify extended emission after subtracting a 
    disk model, \vc{including} a large-scale spiral-like feature. We interpret its signal in the direct vicinity of planet $c$ as tracing the accretion stream that feeds its circumplanetary disk, while the outer part of the feature may rather reflect asymmetric illumination 
    of the outer disk. 
    We also report a bright signal that is consistent with a previously proposed protoplanet candidate enshrouded in dust \vc{near} the 1:2:4 mean-motion resonance with planets $b$ and $c$. 
    The 1.87\,$\upmu$m flux of planet $b$ is consistent with atmospheric model predictions, 
    but the flux of planet $c$ is not. We discuss potential origins for this discrepancy, including 
    significant Pa-$\alpha$ line emission.     
    The 4.83\,$\upmu$m fluxes of planets $b$ and $c$ 
    suggest 
    enshrouding dust or heated CO emission from their circumplanetary environment. 
    }
   {The use of image-processing methods that are optimized for extended disk signals on high-sensitivity and high-stability from 
   \vc{JWST}
   can uniquely identify signatures of planet--disk interactions and enable accurate photometry of protoplanets at wavelengths that are difficult to probe from the ground. Our results indicate that more protoplanets can be identified and characterized in other JWST datasets.}
   
    \maketitle
    
\defcitealias{Mesa2019a}{M19}

\section{Introduction}
    The direct detection and characterization of protoplanets is an emerging research field that is enabled by the latest generations of instruments \vc{achieving} high contrast and high angular resolution that are available on the largest ground-based facilities \citep[e.g.,][]{Keppler2018, Hammond2023}. 
    While the application of post-processing algorithms designed for point-source detection to bright protoplanetary disks has led to a number of unconfirmed protoplanet candidates, the recent development of both iterative and inverse-problem image-processing approaches (IIPAs) is opening the way to unbiased high-contrast IR imaging of the birth environment 
    of planets \citep[e.g.,][]{Pairet2021,Flasseur2021,
    Juillard2023}. 
    \vc{JWST} now offers the opportunity to use these optimized tools to deepen the search and characterization of protoplanets at wavelengths that are inaccessible or are difficult to observe from the ground. This Letter demonstrates the combined potential of JWST and IIPAs and 
    uses 
    the PDS~70 
    system as a testbed.

    \object{PDS 70} is a young ($5.4\pm1$~Myr) K7IV star located at a distance of $113.4\pm0.5~$pc \citep{Muller2018,Gaia-Collaboration2021}. It is surrounded by a protoplanetary disk composed of a 
    water-rich inner disk \citep{
    Dong2012, Perotti2023} that is separated from the outer disk by an annular gap extending up to a radius of 
    $\sim$54~au \citep{Long2018,Keppler2019}. 
   Two nascent planets were imaged and confirmed independently in this large 
   gap, at multiple near-IR (NIR) and submillimeter (submm) wavelengths, as well as in the H$\alpha$ line filter \citep[e.g.,][]{Keppler2018,Muller2018,Christiaens2019, Haffert2019, Isella2019, Benisty2021}. 
   A third protoplanet candidate was also reported 
   on a potential $\sim$13.5\,au orbit 
   \citep[][hereafter \citetalias{Mesa2019a}]{Mesa2019a}.
   The system is therefore an ideal laboratory in which to study planet-disk interactions and search for accretion signatures. Hydrodynamical simulations suggest that the large gap is dynamically carved by both planets, and the (near) 2:1 mean-motion resonance observed by \vc{GRAVITY} \citep{Wang2021} 
   is explained as the outcome of planet migration followed by resonance capture \citep{Bae2019, Toci2020}. 
   Constraints on the distribution and depletion 
   of dust and gas in the disk were recently inferred through radiative transfer and thermo-chemical \vc{forward modeling} of the NIR and submm  observations \citep{Portilla-Revelo2022,Portilla-Revelo2023}. 
   Submm continuum and NIR observations also 
   identified an arm-like structure that is hypothesized to either trace an asymmetric outer ring or a gap-induced 
   vortex \citep[][]{Isella2019, 
   Juillard2022}. 
    
   The spectral characterization of protoplanets $b$ and $c$ favors dust-enshrouded atmospheres combined with potential IR excess \citep{Muller2018, Christiaens2019a, 
   Wang2020, Stolker2020, Wang2021}. 
   The cross-correlation of either these best-fit models or molecular templates with medium-resolution spectra did not detect the planets, however
   \citep{Cugno2021}. This suggests that either significantly more extinction affects the two protoplanets than fitting of the spectral energy distribution (SED) implies or that the self-consistent atmospheric models 
   that have been used to characterize adolescent substellar objects do not describe embedded protoplanets. Hybrid models with contributions from surface accretion shocks \citep[e.g.,][]{Aoyama2020, Aoyama2021}, 
   or circumplanetary disk models may more accurately account for the measured SED \citep[e.g.,][]{
   Chen2022}. 
   In this context, observations at additional wavelengths are key to distinguish the different models. 

   

\begin{table} 
\begin{center}
\caption{Properties of the star(+disk), protoplanets $b$ and c, and candidate $d$ inferred from the NIRCam F187N and F480M observations.} 
\label{tab:results}
\begin{tabular}{lcc}
\hline
\hline
Parameter & F187N & F480M \\
\hline
\multicolumn{3}{c}{PDS~70}\\
\hline
Flux (mJy) & $310.0 \pm 3.1$ & $160.9 \pm 1.6$ \\
\hline
\multicolumn{3}{c}{PDS~70~$b$}\\
\hline
Sep.\ (mas)$^{\rm (a)}$ & $164.5 \pm 10.4$ & $145.6 \pm 20.4$ \\
PA ($\degr$)$^{\rm (b)}$ & $129.0 \pm 4.1$ & $130.4 \pm 6.7$ \\
Contrast$^{\rm (c)}$ & $(2.89 \pm 0.78) \times 10^{-4}$ & $(3.44 \pm 1.50) \times 10^{-3}$\\ 
Flux ($\upmu$Jy) & $89.6 \pm 24.2$ & $ 553.5 \pm 
241.4$\\
\hline
\multicolumn{3}{c}{PDS~70~$c$}\\
\hline
Sep.\ (mas)$^{\rm (a)}$ & $195.7 \pm 23.3 $ & $216.7 \pm 20.6$\\
PA ($\degr$)$^{\rm (b)}$ & $269.7 \pm 6.8$ & $272.8 \pm 3.6$ \\  
Contrast$^{\rm (c)}$ & $(1.37 \pm 0.42) 
\times 10^{-4}$ & $(1.47 \pm 0.30) \times 10^{-3}$\\
Flux ($\upmu$Jy) & $42.5 \pm 13.0$ 
& $ 236.5 \pm 48.3$\\
\hline
\multicolumn{3}{c}{PDS~70~$d$?}\\
\hline
Sep.\ (mas)$^{\rm (a)}$ & $103.4 \pm 23.2 $ & -- \\
PA ($\degr$)$^{\rm (b)}$ & $293.0 \pm 12.7$ & -- \\  
Contrast$^{\rm (c)}$ & $(1.88 \pm 0.69) \times 10^{-4}$ & -- \\
Flux ($\upmu$Jy) & $61.5 \pm 22.6$ & -- \\
\hline
\end{tabular}
\end{center}\addvspace{-0.7em}
{\small%
Notes: 
$^{\rm (a)}$Radial separation. $^{\rm (b)}$Position angle measured east of north. $^{\rm (c)}$Contrast ratio with respect to the star and unresolved inner disk. 
}
\end{table}
    
\section{Observations and image processing} \label{sec:obs}
    \label{sec:image_proc}

       \begin{figure*}[!t]
    \centering
    \includegraphics[width=\textwidth]{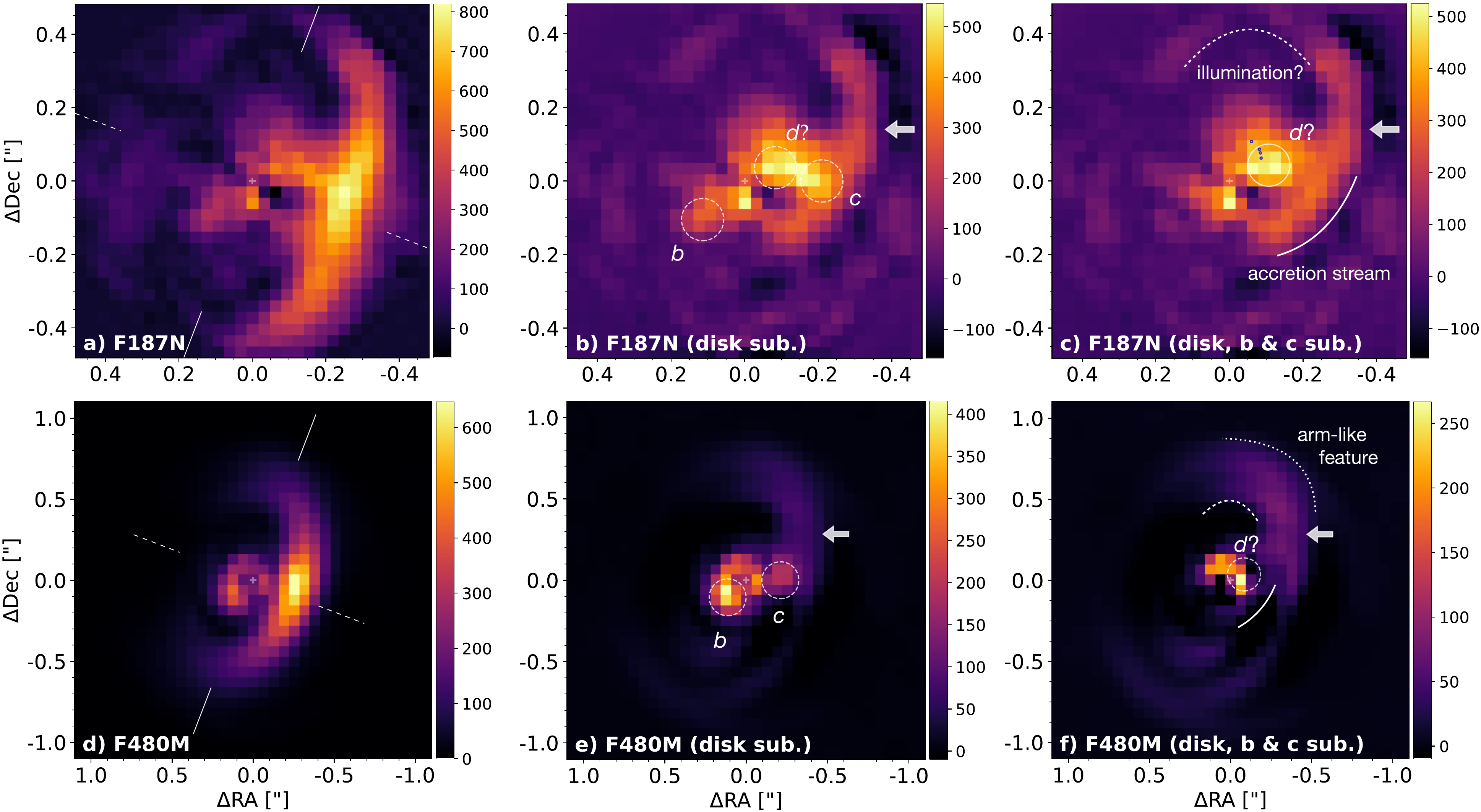}
    \caption{
    NIRCam images of PDS 70 obtained in the F187N (top row) and F480M (bottom row) filters 
    using our iterative PCA algorithm. 
    The second and third columns show the images obtained after subtraction of our outer-disk model 
    and after further subtraction of protoplanets $b$ and c, respectively. The major and minor axes of the disk are indicated in the first column with solid and dashed lines, respectively. The dashed circles indicate the predicted locations of protoplanets $b$ and $c$ based on the orbital fits in \citet{Wang2021} and the location of candidate $d$ based on the orbit suggested in \citetalias{Mesa2019a}. The astrometric measurements for $d$ (blue dots) are compared to our new estimated astrometry (solid circle) in panel c. 
    The units are MJy sr$^{-1}$. 
    }
    \label{fig:FinalImages}
    \end{figure*}
    




    We observed PDS 70 with JWST/NIRCam \citep{Rieke2005} as part of the MIRI guaranteed time observations on protoplanetary disks in the MINDS survey (PI: Th.~Henning, PID: 1282) 
    on 8 March 2023. The images were acquired simultaneously in the F187N ($\lambda_{\rm pivot}=1.874$\,$\upmu$m, $\Delta \lambda=0.024$\,$\upmu$m) 
    and F480M ($\lambda_{\rm pivot}=4.834$\,$\upmu$m, $\Delta \lambda=0.303$\,$\upmu$m) 
    filters without the coronagraph, with pixel scales of 31 and 63 mas/pixel, respectively. 
    We used the smallest subarray (SUB64P), with an effective integration time of 0.35112~s, seven groups per integration, 142 integrations per exposure, and five dithered exposures at two
    roll angles each, separated by $\sim$5.0\degr. The total effective integration time was thus $\sim$8.3 min. 
    
    We calibrated 
    the data with stages 1 and 2 (Detector1 and Image2) of the JWST pipeline \citep[v1.10.2;][]{Bushouse2023}, using the calibration reference data system context \texttt{jwst\_1166.pmap}.
    The calibration 
    yielded two sets of five dithered images, 
    each set corresponding to a different roll angle.
    We then relied on routines from the Vortex Image Processing \citep[VIP;][]{GomezGonzalez2017,Christiaens2023} package
    for image preprocessing (details in Appendix~\ref{sec:preproc}), stellar PSF subtraction, and \vc{forward modeling} of circumstellar signals.
    As the star did not saturate \new{the detector}, 
    we directly performed aperture photometry on the preprocessed images. We set the largest aperture radius that fit within the SUB64P field of view. 
    Table~\ref{tab:results} reports the integrated photometry of the system. We conservatively 
    considered 1\% absolute flux calibration uncertainties. 

    

    We investigated different approaches for PSF modeling and subtraction (details in Appendix~\ref{sec:AltAlgos}). 
    The best-quality images of the system \new{were obtained with} our implementation of an iterative principal component analysis \citep[IPCA; e.g.,][]{Pairet2021, Stapper2022} that leverages roll angle diversity. Its principle relies on estimating the circumstellar signals in the processed image that are obtained at each iteration and removing them from the 
    images that are used to create a PSF model (i.e., for each image, those obtained at the other roll angle) in the next iteration.
    \new{We applied IPCA to a test dataset to illustrate that it reliably recovers point-like and extended circumstellar signals (Appendix~\ref{sec:TestIPCA}).}

    Before the photometry of the protoplanets is extracted, the expected contribution of the disk needs to be removed. This is particularly relevant for planet c, which is located near the bright edge of the outer disk. 
    We relied on the latest radiative transfer models of the disk 
    detailed in \citet{Portilla-Revelo2022, Portilla-Revelo2023}, 
    and followed a procedure that we refer to as the negative fake disk technique (NEGFD; details in Appendix~\ref{sec:NEGFD}).
    After removing the disk contribution, we used the negative fake companion \citep[NEGFC; e.g.,][]{Lagrange2010} technique to extract the exact astrometry and photometry of the protoplanets in both datasets (details in Appendix~\ref{sec:NEGFC_optim}).
    The final astrometry and photometry we retrieved for the protoplanets 
    are presented in Table~\ref{tab:results}. 

\section{Results and discussion}\label{sec:results}

\subsection{Spiral accretion stream or variable illumination?}\label{sec:spiral}

Panels a and e in Figure~\ref{fig:FinalImages} show the F187N and F480M images we obtained with IPCA, respectively. IPCA recovered faint circumstellar signals that were originally hidden in the wings of the stellar PSF while it iteratively corrected for geometric biases that affect extended signals in roll-subtracted images \citep[Fig.~\ref{fig:AltAlgos}a and e; 
    ][]{Juillard2022}. 
The IPCA images 
reveal signals of the inner and outer disk, signals at the expected location of the protoplanets, and a shift in the maximum intensity of the outer disk north of the semiminor axis $\beta$ on the near side of the disk (located at a position angle east of north PA$_{\beta}$$\sim$$249\degr$). In the absence of disk asymmetry or protoplanets, the maximum in total intensity is otherwise expected to be located along PA$_{\beta}$, based on the scattering phase function of sub-$\upmu$m size dust grains \citep[e.g.,][]{Milli2017}.
To 
highlight 
the protoplanets and any disk asymmetry, we subtracted the 
disk model found with NEGFD from the IPCA images (Fig.~\ref{fig:FinalImages}c and f). This revealed residual extended signals in addition to protoplanets $b$ and c, the predicted locations of which are indicated for the epoch of the observations. These predictions are based on the orbital fits presented in \citet{Wang2021}, and they are available through the platform \texttt{whereistheplanet} 
\citep{Wang2021b}. 
We do not detect any significant counterpart for the proposed submm continuum signal at the L5 Lagrangian point associated with planet $b$ \citep{Balsalobre-Ruza2023}.
We note an outstanding extended spiral-like signal connected with the position of planet $c$, however. It is indicated with a thick arrow in Fig.~\ref{fig:FinalImages}b and e.

\new{The dynamical interaction between a protoplanet and the disk in which it is embedded has long been known to cause spiral density waves \citep[][]{ Goldreich1979,Ogilvie2002} and angular momentum transport \citep[][]{Lin1979, Rafikov2002}. In the vicinity of the planet, the latter is expected to lead to a spiral-shaped accretion stream \citep[][]{Lubow1999, Ayliffe2009}. The 
accretion stream associated with 
PDS~70 $c$ was predicted in a dedicated 3D hydrodynamical simulation in \citet{Toci2020}. Figure~\ref{fig:SpiralTrace} compares it with the spiral-like signal we identified in our observations.}
Although the agreement is remarkable, we note that the accretion stream mostly delimits the edge of the cavity. 
An asymmetric 
illumination of the outer-disk edge that is not captured by our radiative transfer disk model could equally lead to an excess signal in our images after disk model subtraction. 
Multi-epoch polarized intensity images of the system show varying illumination and shadowing effects 
in the northwest and southeast parts of the outer disk \citep[e.g., Fig.~A2 in][]{Juillard2022}, suggesting that this is likely the dominant cause for the observed excess. 
This interpretation is consistent with the significant mid-IR (MIR) SED variability measured for the system, \vc{which} also suggests variable shadowing effects from the inner parts onto the outer parts of the disk \citep{Perotti2023}.

\begin{figure*}[!t]
\centering
\includegraphics[width=\textwidth]{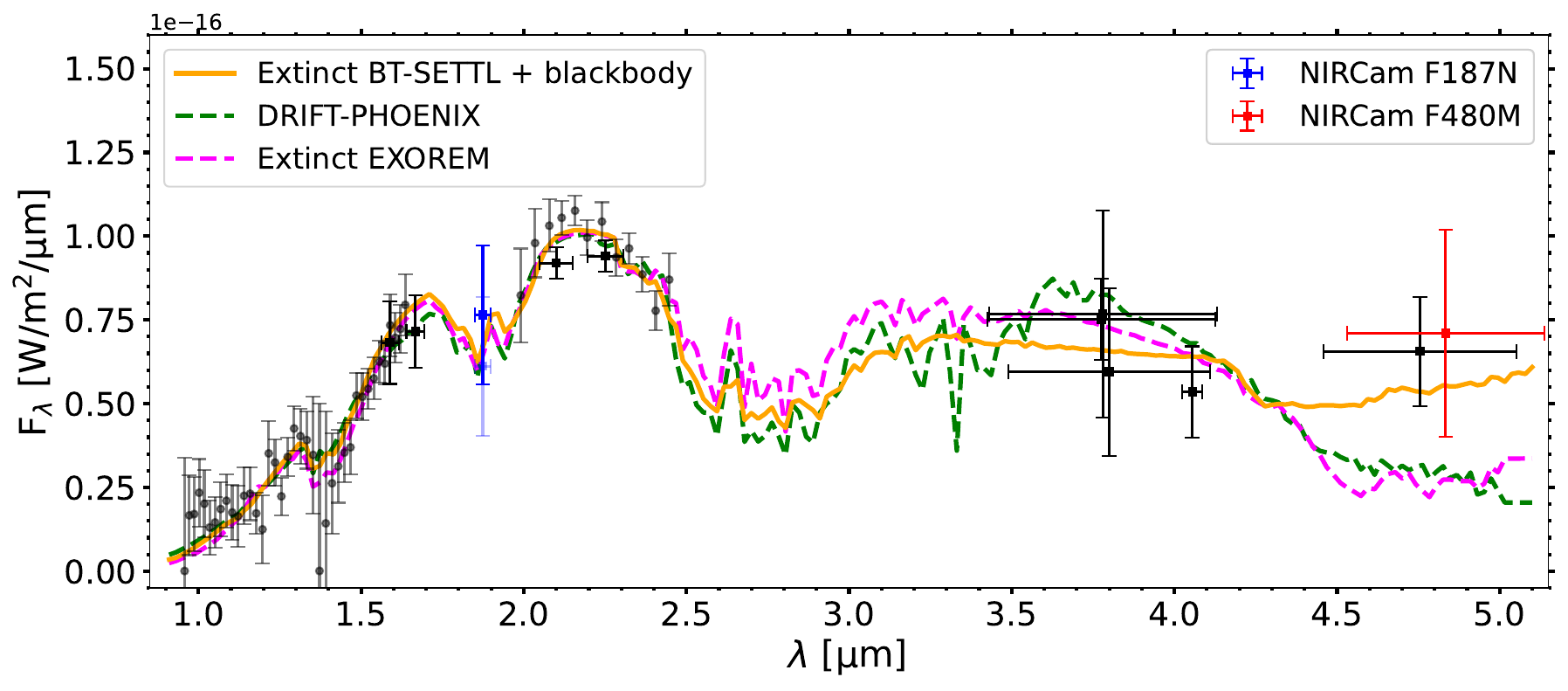}
\caption{Composite SED of PDS 70 $b$ showing spectro- and photometric measurements from the literature (gray and black error bars, respectively), the new NIRCam F187N and F480M photometry 
(blue and red error bars, respectively), and 
the best-fit atmospheric models found in \cite{Wang2021}. The model with the most support is shown with a solid orange line (extinct BT-SETTL model with additional blackbody emission). It is consistent with both of our measurements, and suggests that circumplanetary contribution is required. 
The light blue error bar is obtained 
considering photometry from the literature for the star. 
It illustrates the uncertainty associated with variability that affects \new{some} other data points of the SED (details in Appendix~\ref{sec:F187N_excess}). 
}
\label{fig:spec_b}
\end{figure*}

While illumination effects most likely cause excess signals at the edge of the outer disk (upper arc in Fig.~\ref{fig:FinalImages}c), these effects alone 
cannot account for the part of the spiral-like signal that is located inside the cavity based on the level of dust depletion therein 
\citep[e.g.,][]{Dong2012, Keppler2018}. In the direct vicinity of planet $c$ (lower arc), a genuine dust density enhancement therefore appears to be required. Based on the clear visual connection to the location of protoplanet $c$, a spiral accretion stream feeding the circumplanetary disk of planet c that is detected in submm continuum observations \citep{Isella2019, Benisty2021, Casassus2022} is the most straightforward explanation for this signal.
Excess signal has tentatively been observed there in IIPA-processed VLT/SPHERE images  
\citep[][]{Flasseur2021, Juillard2022}, although the nonremoval of 
an outer-disk model complicates an unambiguous identification of this signal in these images. 
Moreover, the location of the accretion stream is coincident with a 
gap-crossing spur found in ALMA CO observations of the disk \citep{Keppler2019}. This further strengthens the interpretation that this is an accretion stream. 
This result suggests that some of the observed spiral features in less strongly inclined disks than PDS~70 could \new{also} be associated with embedded protoplanets \new{\citep[e.g.,][]{Dong2018, Ren2024}}. A similar spiral-shaped signal has recently been identified as connected to HD~169142~$b$ \citep{Hammond2023}. Likewise, a gap-crossing filament coincident with a twist in one of the 
spirals of HD~135344~B may also \new{be caused by} an embedded protoplanet \citep{Casassus2021}.

An arm-like feature was 
identified and characterized in the outer disk of PDS~70 using multi-epoch SPHERE images of the system obtained with IIPAs \citep{Pairet2021, Juillard2022}. The trace of this arm in a 2021 SPHERE dataset is compared with our IPCA image in Fig.~\ref{fig:SpiralTrace}c.
It is unclear whether it is associated with the spiral accretion stream 
or traces a separate feature, such as a vortex or an asymmetric second ring \citep[][]{Juillard2022}. 
The inner part of the arm at the edge of the cavity appears to be consistent with the 
part of the spiral-like signal seen in the F187N image, which may trace the varying illumination of 
the outer-disk edge. The outer part of the arm appears to be consistent with the arm-like feature identified in previous SPHERE images. 
This 
feature 
is the only signal 
that is detected at a signal-to-noise ratio (S/N) $> 5$ in our F480M images in addition to the protoplanets (Fig.~\ref{fig:SNRmaps}) and was also observed in Keck/NIRC2 images of the system after a similar NEGFD procedure as we used \citep[Fig.~1 in][]{Wang2020}. 
An alternative origin for the arm-like signal is the presence of an as yet undetected planet in the outer disk that excites an inner spiral wake. We therefore investigated whether our data are sensitive to additional planets (Appendix~\ref{sec:contrast_curves}).
Our F480M contrast 
and corresponding mass-sensitivity curves (Fig.~\ref{fig:contrast_mass_limits}) 
constrain any 
planets in the outer disk to have a mass below $\sim$1--2 $M_J$, neglecting extinction.


\subsection[Astrometry and photometry of protoplanets b and c]{Astrometry and photometry of protoplanets {\sf b} and {\sf c}}\label{sec:photometry}

The astrometry and contrast of planets $b$ and $c$ with respect to the star were inferred by using 
our NEGFC approach (Appendix~\ref{sec:NEGFC_optim}) after subtraction of our optimal disk model (Appendix~\ref{sec:NEGFD}). These contrast values were then multiplied by the integrated stellar flux values reported in Table~\ref{tab:results} to obtain the flux of the protoplanets. We note that the stellar fluxes in the F187N and F480M filters are $\sim$25\% and $\sim$7\% brighter than estimated based on the 
SpeX spectrum presented in \citet{Long2018} and the best-fit SED model to the extracted MIRI-MRS spectrum \citep[][
]{Perotti2023}, respectively. 
These differences are 
compatible with the significant 
IR variability of the star and inner-disk rim 
\citep[e.g., $\sim$25\% variation at $\sim$5\,$\upmu$m between Spitzer and JWST observations;
][]{Perotti2023}. 
These considerations inspire caution regarding protoplanet photometry derived in contrast of the star \vc{that are} based on nonconcurrent absolute star photometry measurements. 

\begin{figure*}[!t]
\centering
\includegraphics[width=\textwidth]{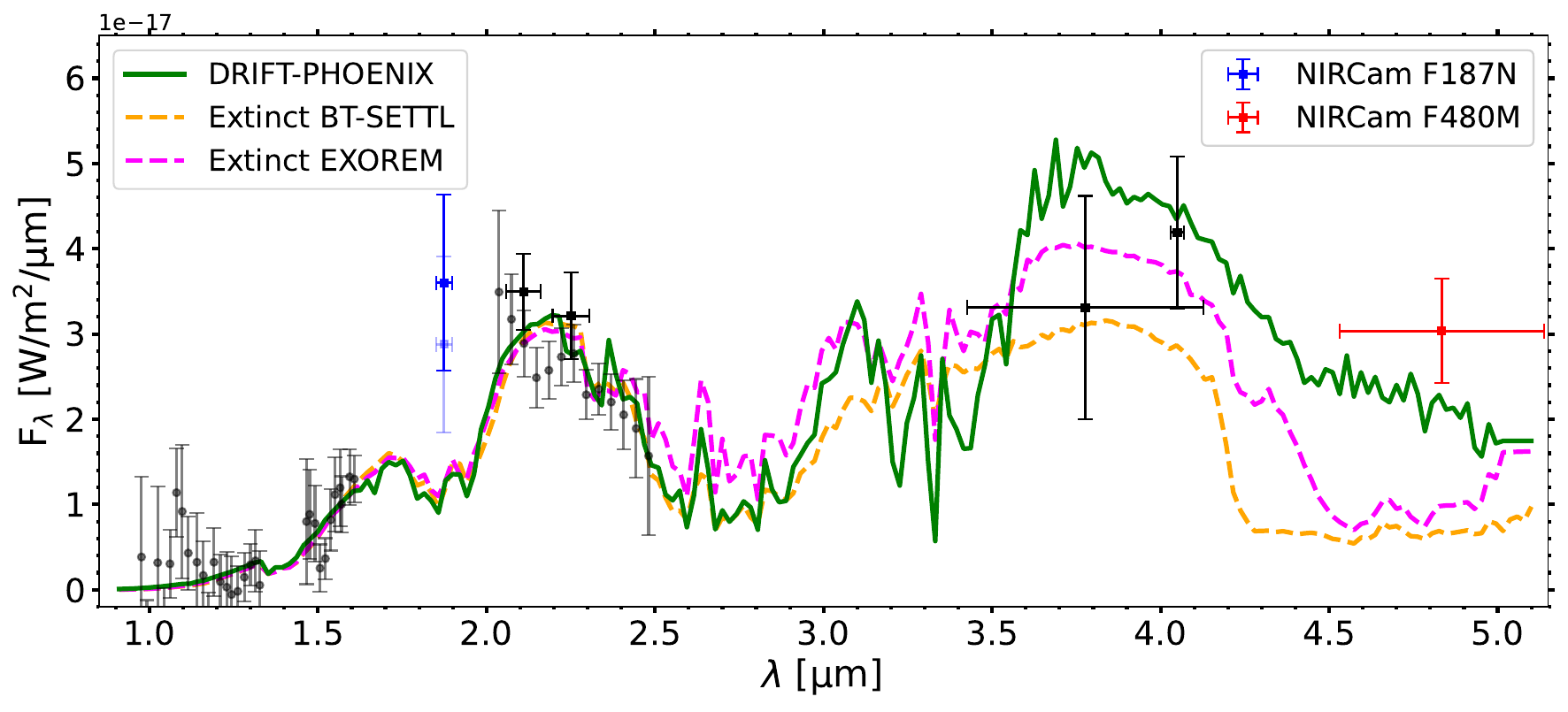}
\caption{Same as Fig.~\ref{fig:spec_b}, but for PDS~70~$c$. 
Here, the model with the most support is the plain Drift-Phoenix model \citep[details in][]{Wang2021}. 
}
\label{fig:spec_c}
\end{figure*}

For each planet, the astrometric values we found for the two filters are consistent with each other. The F187N astrometry of planet 
$c$ and candidate $d$ is affected by large uncertainties owing to neighboring disk signals and the overlapping spiral accretion stream. For the F480M images, the large uncertainties also reflect the coarser angular resolution. All estimates are consistent with the expected astrometry of the two protoplanets at the date of our observations (given in the last column of Table~\ref{tab:comparison_NEGFC}), based on the orbital fits presented in \citet{Wang2021}. 
Considering our large uncertainties compared to ground-based measurements, we do not attempt new orbital fits 
in this work. 


Figures~\ref{fig:spec_b} and \ref{fig:spec_c} show the 
SED of PDS 70 $b$ and $c$, respectively, including our NIRCam data at 1.87 and 4.83\,$\upmu$m. 
The F187N measurement for planet $b$ is consistent with predictions from the best-fit atmospheric models obtained with the BT-Settl, Drift-Phoenix, and EXOREM grids 
\citep[details in][]{Wang2021}. This does not hold for planet $c$. 
This discrepancy can have various (not mutually exclusive) causes that we discuss in Appendix~\ref{sec:F187N_excess}. 
Here, we discuss 
the hypothesis of significant Pa-$\alpha$ line emission from the protoplanet that is not captured in the atmospheric model. 
If the excess signal above the model atmospheric contribution comes from Pa-$\alpha$ emission alone, the line flux is
$(5.7 \pm 2.7) \times 10^{-19}$ W m$^{-2}$.
Given the unknown amplitude of other biases discussed in Appendix~\ref{sec:F187N_excess}, this estimate is a conservative upper limit.
Considering $A_V \approx 2.0$ mag \citep{Uyama2021} and the distance of the system, our constraint on the Pa-$\alpha$ luminosity is  $\log(L_{{\rm Pa}\alpha, c}/L_{\odot}) \lesssim -6.5^{+0.2}_{-0.3}$, where $L_{\odot}$ is the solar luminosity. 
\vc{When} a planet mass of $M_c \sim 7 M_J$ \citep{Wang2021} is assumed, the models in \citet{Aoyama2021} suggest that the mass accretion rate would be $ \log(\dot{M_c}/M_J$ yr$^{-1}) \lesssim -6.7^{+0.3}_{-0.2}$.
The same assumptions for the model and the planet applied to the H$\alpha$ flux reported in
\citet{Haffert2019} 
yield an accretion rate
$\log(\dot{M_c}/M_J$ yr$^{-1}) \approx -7.7$. If the \citet{Aoyama2021} models are an accurate representation of the accretion process onto giant planets, our results suggest that either additional sources of bias may have a non-negligible effect (e.g., stellar variability or underestimated SPHERE/IFS measurements; Appendix~\ref{sec:F187N_excess}), or
that protoplanets undergo significant accretion variability 
\citep[e.g.][]{Szulagyi2020, Casassus2022}. 
With a similar reasoning as above, but using the 3 $\sigma$ uncertainty on the F187N flux measured for planet $b$ and an assumed extinction of $A_V \sim 0.9$ \citep{Uyama2021}, we constrain the Pa-$\alpha$ luminosity to $\log(L_{{\rm Pa}\alpha, b}/L_{\odot}) < -6.2$. 
Based on the planet accretion models in 
\citet{Aoyama2021}, this translates into an 
upper mass-accretion rate limit of 
$\log(\dot{M_b}/M_J$ yr$^{-1}) < -6.0$
for a planet with a mass $M_b \sim 3 M_J$ 
\citep{Wang2021}. 
This constraint is in line with the accretion rates inferred from the H$\alpha$ flux measured for $b$ \citep{Haffert2019}, which for the same model and assumptions leads to 
$\log(\dot{M_b}/M_J$ yr$^{-1}) \approx -7.5$.
In summary, the measurement uncertainties and additional sources of bias together prevent us from confirming significant Pa-$\alpha$ line emission for the two protoplanets.

Our new F480M measurement for PDS~70~$b$ is consistent with the NACO $M$-band measurement presented in \citet{Stolker2020}. \citet{Wang2021} found that this point was driving the inclusion of an additional blackbody contribution that is representative of heated circumplanetary dust in the atmospheric model with the most support (solid line in Fig.~\ref{fig:spec_b}). 
For PDS~70~$c$, 
our 4.8 $\upmu$m photometry is roughly compatible with the best-fit Drift-Phoenix model, but it is significantly higher than the best-fit models from the other two grids, which are known to reproduce the spectra of old L-T dwarfs better \citep[e.g.,][]{Witte2011}.
For both planets, the tentative excess may be attributable to either a warm dusty environment or to ro-vibrational CO line emission from a heated circumplanetary disk \citep[e.g.,][]{Oberg2023}.
We defer a detailed spectroscopic analysis to a later study including NIRSpec measurements. The NIRSpec measurements 
have the highest potential to confirm the tentative 4.8 $\upmu$m excesses and constrain their origin.

\subsection{A third protoplanet, a dust clump, or an inner spiral?}

The brightest signals in our disk-subtracted F187N image (Fig.~\ref{fig:FinalImages}b) are found near the predicted location of a protoplanet candidate proposed in \citetalias{Mesa2019a} at the outer edge of the inner disk (referred to as a point-like feature therein). The candidate was found in SPHERE/IFS datasets acquired between May 2015 and April 2019. Its reported astrometry, indicated with blue dots in Fig.~\ref{fig:FinalImages}, is consistent with an $\sim$13.5\,au circular Keplerian orbit in a plane similar to that of the outer disk, which 
is assumed for the predicted location shown with dashed circles in Fig.~\ref{fig:FinalImages}. 
For clarity, we also show the images that were obtained after the estimated flux of protoplanets $b$ and $c$ was also subtracted (Fig.~\ref{fig:FinalImages}c and f).
While inner-disk signals are present near the predicted location, there appears to be a significant excess compared to the disk signals alone, considering that 
the inner-disk emission originates in a symmetric location with respect to the minor axis 
southeast of the star. 
While a signal at a separation of $\sim$$2\lambda/D$ from the star should be considered with caution, the recovery of the inner disk with a geometry close to expected suggests that the PSF subtraction residuals are lower than the inner-disk signals. 
This means that the observed excess is likely 
of 
circumstellar origin and not an 
artifact.
We therefore refer to it hereafter as ``$d$'' because it might trace a dusty feature or a third protoplanet. 

Table~\ref{tab:results} reports the astrometry and photometry we extracted for $d$ in the F187N image. While the F480M image also shows a bright pixel near the expected location, its separation from the star is too small 
for a reliable contrast estimate. 
The contrast of $d$ derived in the F187N image is higher about a factor of 2 than the median contrast reported in $YJH$ bands in \citetalias{Mesa2019a}. Our estimate is affected by large uncertainties due to the unknown amount of contamination from the inner disk, and this excess is therefore only marginally significant. Nonetheless, we argue that this excess 
is expected. The source spectrum is dominated by scattered stellar light at NIR wavelengths \citepalias{Mesa2019a}, which is consistent with a very dusty object. In this case, the scattering phase function of the total intensity is also expected to modulate its brightness along its orbit. Compared to prior epochs, the object is now closer to PA$_{\beta}$, and therefore, we would  
expect a 
brighter signal. We estimate an enhancement of a factor of $\sim$1.6 in reflected light for $d$ between April 2019 and March 2023 based on the variation in the flux of the outer disk measured at PA=PA$_{\beta}$-PA$_{d,2019}$ and PA=PA$_{\beta}$-PA$_{d,2023}$, where we considered the southwest part of the outer disk for this estimate to avoid any bias from the accretion stream toward the northwest. 
Within the uncertainties, our measured contrast is therefore consistent with tracing the same object as \citetalias{Mesa2019a}.
This 
new measurement adds 4 years 
to the existing 4-year time baseline for the orbital coverage, and it significantly reduces the probability that $d$ either traces a moving illumination effect or the filtered northwestern tip of the inner disk (in \citetalias{Mesa2019a}). 


\new{Our independent redetection of a signal that is compatible with candidate $d$ does not unambiguously confirm its protoplanet nature, but}
it raises the question of which other physical processes might give rise to a bright and compact NIR signal that moves at \vc{the} local Keplerian speed. 
The $YJH$ spectrum measured in \citetalias{Mesa2019a} is consistent with tracing scattered stellar light, and the authors therefore suggested that it might trace either a transient dust clump 
or a protoplanet enshrouded in dust. 
Radiative hydrodynamical simulations of embedded giant planets suggest that they become enshrouded in a dusty circumplanetary disk or envelope, which can display a scattered-light dominated spectrum at NIR wavelengths \citep[e.g.,][]{Szulagyi2019}.
If $d$ indeed traces a protoplanet enshrouded in dust, its semimajor axis of $\sim$13.5\,au would place it near the 1:2:4 mean-motion resonance with planets $b$ and $c$ 
\citep[$a_b\approx 21$\,au and $a_c \approx 34$\,au;][]{Wang2021}. Follow-up studies of $d$
are therefore especially exciting. 
Distinguishing the hypotheses of a 
dust clump from the circumplanetary disk or an envelope will require MIR flux measurements \citep[e.g.,][]{Chen2022}. Because of the angular separation, this may need the advent of MIR imagers and spectrographs on extremely large telescopes \citep[e.g., ELT/METIS;][]{Brandl2018}. Figure~\ref{fig:schema} summarizes our proposed interpretation of the main features detected in our NIRCam observations of PDS~70.
\begin{acknowledgements}
We thank Jason Wang for sharing atmospheric models and GRAVITY spectra of the protoplanets. We also thank Yuhiko Aoyama, Faustine Cantalloube and Julien Girard for useful discussions. VC and OA thank the Belgian F.R.S.-FNRS, and the Belgian Federal Science Policy Office (BELSPO) for the provision of financial support in the framework of the PRODEX Programme of the European Space Agency (ESA) under contract number 4000142531. This project has received funding from the European Research Council (ERC) under the European Union's Horizon 2020 research and innovation programme (grant agreement No 819155), and from the Wallonia--Brussels Federation (grant for Concerted Research Actions).
G-DM acknowledges the support of the DFG priority program SPP 1992 ``Exploring the Diversity of Extrasolar Planets'' (MA~9185/1) and from the Swiss National Science Foundation under grant
200021\_204847
``PlanetsInTime''. Parts of this work have been carried out within the framework of the NCCR PlanetS supported by the Swiss National Science Foundation.
TPR acknowledges support from the ERC under grant 743029 (EASY).
This work is based on observations made with the NASA/ESA/CSA James Webb Space Telescope. The data were obtained from the Mikulski Archive for Space Telescopes at the Space Telescope Science Institute, which is operated by the Association of Universities for Research in Astronomy, Inc., under NASA contract NAS 5-03127 for JWST. 
This work has made use of data from the European Space Agency (ESA) mission {\it Gaia} (\url{https://www.cosmos.esa.int/gaia}), processed by the {\it Gaia} Data Processing and Analysis Consortium (DPAC, \url{https://www.cosmos.esa.int/web/gaia/dpac/consortium}). Funding for the DPAC has been provided by national institutions, in particular the institutions participating in the {\it Gaia} Multilateral Agreement. 
This work benefited from the 2022 Exoplanet Summer Program in the Other
Worlds Laboratory (OWL) at the University of California, Santa Cruz, a program funded by the Heising-Simons Foundation.
\end{acknowledgements}

\bibliographystyle{aa} 
\bibliography{references}

\clearpage

\begin{appendix}

\section{Image processing}

    \begin{figure*}[h]
    \centering
    \includegraphics[width=\textwidth]{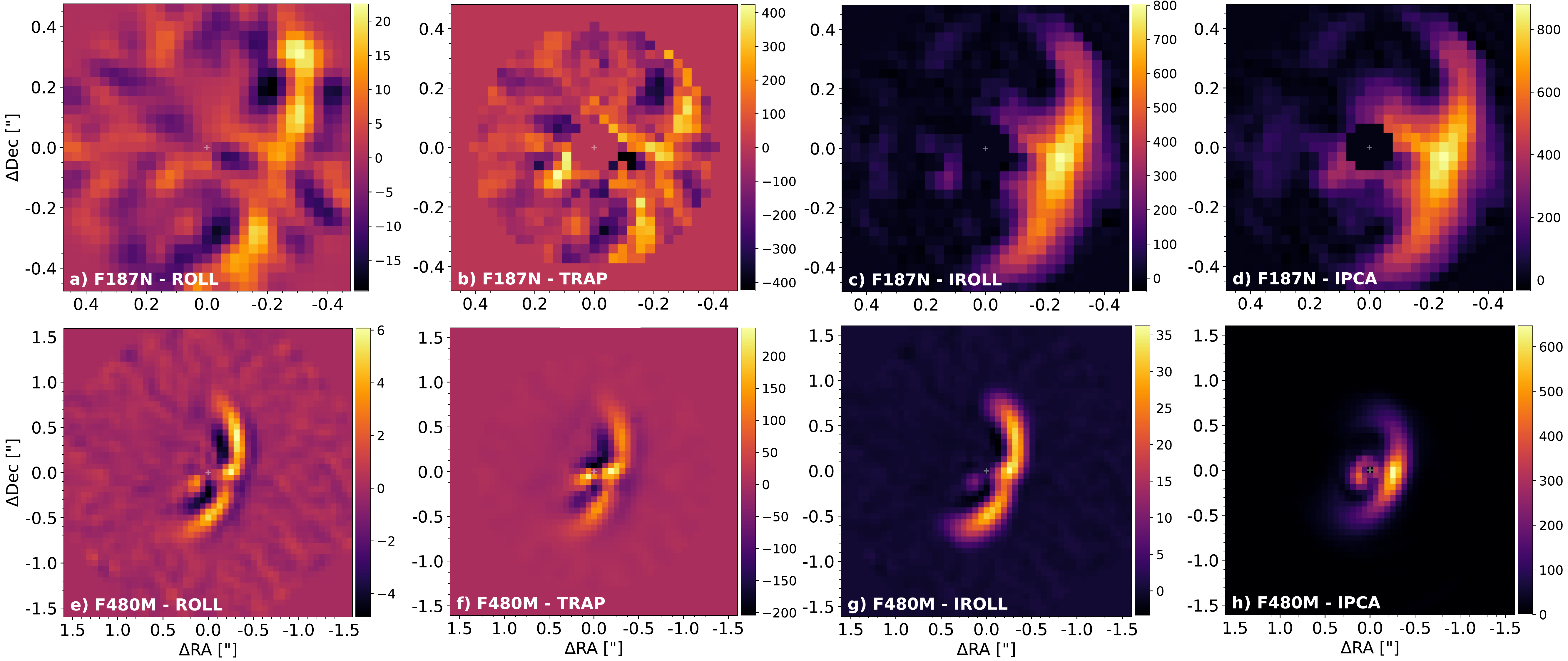}
    \caption{Images obtained at 1.87$\upmu$m (F187N; top row) and 4.80$\upmu$m (F480M; bottom row) with mean roll subtraction, 
    TRAP, IROLL, and IPCA. See text for the details of each algorithm. The images correspond to the largest common field of view probed by the dithering pattern employed during the observation. The plate scale is 31 and 63\,mas/pixel for the F187N and F480M images, respectively. A numerical mask with the radius set to the FWHM of the PSF covers the inner part of the F187N images. 
    }
    \label{fig:AltAlgos}
    \end{figure*}

\subsection{Preprocessing} \label{sec:preproc}

    We first identified remaining bad pixels in the 
    calibrated images through sigma-clipping using the \texttt{cube\_fix\_badpix\_clump} function of VIP, 
    and we corrected for them using Gaussian kernel interpolation. We subsequently corrected for the imperfect background subtraction performed by the JWST pipeline that led to negative values by subtracting the residual background level estimated far from the star.
    We then found the stellar centroid coordinates using the \texttt{cube\_recenter\_dft\_upsampling} routine in VIP. This routine first leverages the single-step discrete Fourier transform algorithm presented in \citet{Guizar-Sicairos2008} to 
    find
    shifts that optimize the image cross-correlation throughout the sequence, and it then fits
    a 2D Gaussian model to the mean image of the aligned cube to find the centroid coordinates. 
    Throughout all stages of the image processing, all shift and rotation operations were performed in the Fourier plane, the default behavior in VIP\footnote{\url{https://github.com/vortex-exoplanet/VIP}}, as this better preserves pixel intensities \citep[e.g.,][]{Larkin1997}. 
    
\subsection{Subtraction of the point spread function}\label{sec:AltAlgos}
    
    We investigated different approaches for the modeling and subtraction of the PSF.
    We first considered pair-wise roll-subtraction \citep[e.g.][]{Schneider2003} between individual dithered images acquired with each of the two roll angles. We also considered the pair-wise subtraction of the mean image of each of the two sets. We note a minor improvement in the residuals near the star that might be due to the spatial undersampling of the images \vc{which limits} the efficacy of individual pair-wise image subtractions. Both the mean and individual options are implemented in the \texttt{roll\_sub} function of VIP, which is available as of version 1.6.0. 
    The images were smoothed with a Gaussian kernel, the FWHM of which was set to 60\% of the observed width of the instrumental PSF. This helps to mitigate the undersampling that affects the F187N data and the pixel-to-pixel noise induced by the roll-subtraction approach. In Fig.~\ref{fig:AltAlgos}a and e, we show the images we obtained with mean roll subtraction on the F187N and F480M data, respectively. Low residuals are achieved owing to the stability of the observed PSF, but significant self-subtraction and geometric distortion of circumstellar signals can also be noted (e.g., by comparison with the IPCA images). Compared to the F480M images shown in Fig.~\ref{fig:FinalImages}, the F480M images in Fig.~\ref{fig:AltAlgos} correspond to the largest common field covered by all dither positions.

In addition to direct roll subtraction, we investigated alternative post-processing methods designed to work in combination with angular differential imaging, namely principal component analysis \citep[PCA;][]{Soummer2012, Amara2012} and temporal reference analysis for planets \citep[TRAP;][]{Samland2021}. We used the PCA algorithm implemented in VIP and used the images obtained at the other roll angle as input PCA library for each image. This is similar to smartPCA, which was proposed for angular differential imaging \citep{Absil2013}. This yielded very similar results to the roll-angle subtraction.
The TRAP algorithm computes a contrast map. A contrast was computed for each off-axis ($\Delta$RA, $\Delta$DEC) pair on-sky by simultaneously modeling the pixel light curves created by an off-axis PSF at the sky position and the temporal systematics of the pixels affected by this off-axis PSF. For our JWST data, there are measurements at two roll angles, such that the off-axis PSF model for each ($\Delta$RA, $\Delta$DEC) pair resulted in a step-function with varying jump heights depending on the location of each pixel relative to the off-axis PSF.
In practice, the contrast map resulting from the \vc{forward modeling} is a form of local deconvolution with a PSF model that takes the field-of-view rotation into account.
This approach is optimized for detecting point sources, but it also works out of the box. However, because the disk structure surrounding the host star is complex, it is difficult to locate uncontaminated reference pixels of the host star PSF to model the temporal systematics in the data. For our results, each off-axis PSF pixel light curve (step function) was fit simultaneously with only a constant factor. Because the instrument is stable, this provided good results. However, higher-order noise models or models informed by temporal trends seen in the guide-star observations may provide better results in the future. Especially when fitting a \vc{forward model} that explicitly models the entire scene including disks and planets, overfitting becomes less of a concern. The TRAP analysis is extremely efficient computationally. It requires less than 5 seconds on a laptop. The TRAP images obtained for the F187N and F480M data are shown in Fig.~\ref{fig:AltAlgos}b and f, respectively.

As a less aggressive alternative, we performed reference star differential imaging \citep[RDI; e.g.,][]{Mawet2012} using 
the PCA algorithm implemented in VIP. 
We tested RDI using observed noncoronagraphic PSFs from the MAST archive as reference library, but this led to strong PSF residuals after subtraction.
It is unclear whether this is due to the field dependence of the PSF, undersampling effects, a different spectral slope for the source (for the F480M filter), or a combination of these factors.  
We therefore also tested creating reference stars using the {\sc webbpsf} package \citep[v1.1.1;][]{Perrin2014} as reference PSFs to test RDI. In the latter case, we considered the PSF distortion dependence on the location in the field. We built the synthetic PSFs considering a 3972~K blackbody \citep{Muller2018} as the stellar spectrum model, anticipating a potential impact of the spectral slope on the PSF for the F480M filter. We also used the optical path differences from the day following the observations, which appear identical to those measured 3 days before the observations. 
Upon carefully checking the radial profile of the {\sc webbpsf} synthetic PSFs and the observed PSF of PDS~70, we noted a broader core for the observed PSF of PDS~70 and residual diffuse hexagonal-shape stellar wings after subtraction. This may suggest a non-negligible contribution from marginally resolved inner-disk signals. We also tested PCA-RDI with a library of synthetic PSFs. These were produced separately for each observed dither position on the detector and considered a range of subpixel shifts around these dither locations (within 0.5px, with steps of 0.05px). Here, we attempted to emulate the broader core and undersampling effects. 
This marginally improved the results, but still yielded strong PSF wing residuals, and hence, poor-quality final images. 

    
As our RDI attempts were unfruitful, 
we instead focused on iterative approaches leveraging roll-angle diversity. 
We implemented an iterative roll (IROLL) subtraction algorithm similar to the one presented in \citet{Heap2000}, and an iterative principal component analysis algorithm \citep[IPCA; e.g.,][]{Pairet2021}. In either case, circumstellar signals estimated in the processed image obtained at each iteration (e.g., positive residuals above a certain threshold) were removed from the pre-processed images used to create a PSF model (i.e., for each image, those obtained at the other roll angle) in the subsequent iteration.
IROLL directly uses the images $B_i$ obtained at roll angle $b$ as PSF model for images $A_i$ obtained at roll angle $a$ (and vice versa for the roles of $A_i$ and $B_i$) and iteratively directly removes the estimated circumstellar component from the roll images.
The difference in the IPCA algorithm is that for images $A_i$, the principal components are learned from images $B_i$ and are then projected onto the $A_i$ images to produce PSF models for subtraction (and vice versa). In practice, we used the first principal component with a cube reduced to two images corresponding to the mean image of each roll-angle observation. Thus, the only difference between IPCA and IROLL resided in the projection of a normalized mean PSF (i.e., the projected first principal component) in the former case.
    
While previous IPCA implementations considered all strictly positive residuals \citep[e.g.,][]{Pairet2021, Juillard2023}, here, we set an absolute threshold slightly above the noise level achieved in the roll-subtraction image, namely 10\,MJy/sr and 1\,MJy/sr in the F187N and F480M data, respectively. An absolute threshold like this should be used with care as it may not be appropriate for all datasets (e.g., images with a strong radial dependence on the residual noise level). We realized that this was relevant given the relatively constant (self-subtracted) noise level in the image. It was therefore particularly efficient at mitigating the propagation of ring-like artifacts \citep[observed in][]{Pairet2021, Juillard2023}.

The estimated circumstellar signal map at each iteration was smoothed using a thin 2D Gaussian kernel set to a one-pixel FWHM before its removal from the roll images that were used to produce the PSF model images at the subsequent iteration. This intends to capture the spatial correlation of neighboring pixels and leads to a better recovery of faint signals that are originally drowned in the noise level of the roll-subtraction image.
Our IPCA algorithm iteratively corrects not only for \textit{self-subtraction} because circumstellar signals are iteratively removed from the image library that is used to calculate the principal components, but also for \textit{oversubtraction}. This is because the model PSF image that is subtracted is built from the projection of principal components onto the original images minus the estimated circumstellar signal map obtained at the previous iteration.

    
We implemented an automatic convergence criterion based on a user-defined relative tolerance (default 1e-4). When all pixel intensities in the image obtained at the subsequent iteration vary by less than this relative tolerance, the algorithm stops and considers to have converged onto a final image.
While this criterion worked for our data, likely helped by the high stability of the PSF, 
we highlight that there is no mathematical guarantee that this fix-point algorithm will systematically converge in general \citep[see also][]{Juillard2023}.
Both IROLL and IPCA converged to a final image within $\sim$1000 iterations. While IROLL recovered a significant fraction of the self-subtracted signals and yielded a similar final image as IPCA in the F187N image, this was not the case for the F480M filter image. In the latter case, it does not appear to converge onto an image free of geometric biases \citep[see e.g.,][]{Juillard2022}.

\begin{figure}[h]
\centering
\includegraphics[width=\columnwidth]{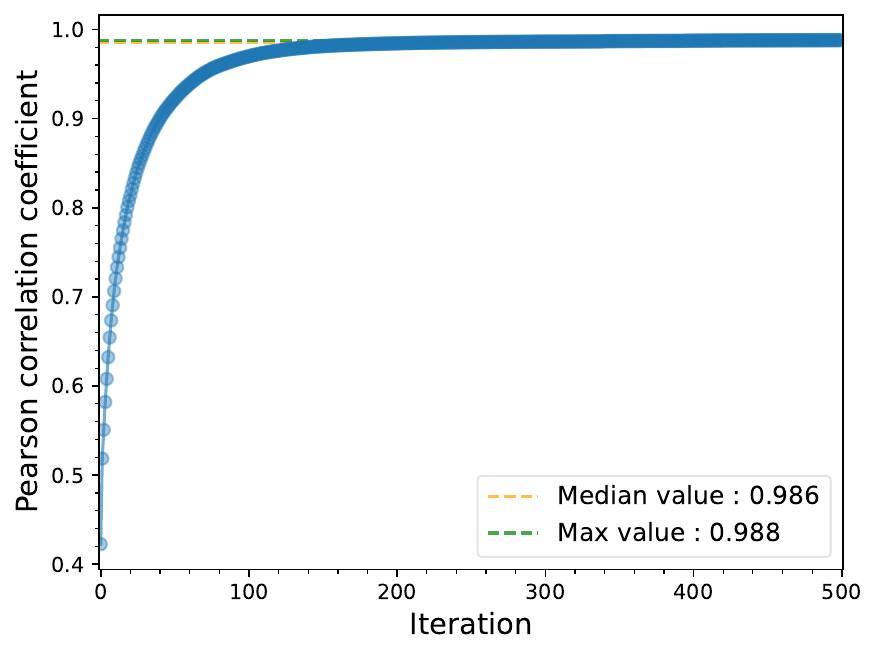}
\caption{Pearson cross-correlation coefficient calculated between the F187N image obtained by IPCA at each iteration and our radiative outer-disk model. Our disk model is not used by IPCA, but rather as a diagnostic for the efficiency of the IPCA algorithm. The geometric biases induced by roll subtraction are corrected within $\sim$200 iterations.}
\label{fig:cross-corr_IPCA_model}
\end{figure}

As a safety check that the algorithm properly recovered the outer disk, we show in Fig.~\ref{fig:cross-corr_IPCA_model} the Pearson cross-correlation calculated between the F187N image obtained at each iteration of IPCA and our corresponding radiative transfer model of the outer disk in a mask encompassing the south part of the outer disk (i.e.,~all the signals south of the peak intensity in Fig.~\ref{fig:disk_models}a; see Sec.~\ref{sec:NEGFD} for more details about the disk model). We highlight that this disk model is not used by IPCA, which functions in an agnostic and automated manner. Most of the geometric biases are corrected within $\sim$200 iterations. Most of the flux is also recovered within a similar number of iterations, with only marginal gains beyond $\sim$200 iterations.

The individual and mean roll subtraction, iterative roll subtraction, and iterative PCA algorithms were all implemented in VIP and are available as of version 1.6.0. The images obtained with roll subtraction, TRAP, iterative roll subtraction, and iterative PCA are shown in Fig.~\ref{fig:AltAlgos}.


\subsection{Reliability of the IPCA}\label{sec:TestIPCA}

    \begin{figure*}[h]
    \centering
    \includegraphics[width=\textwidth]{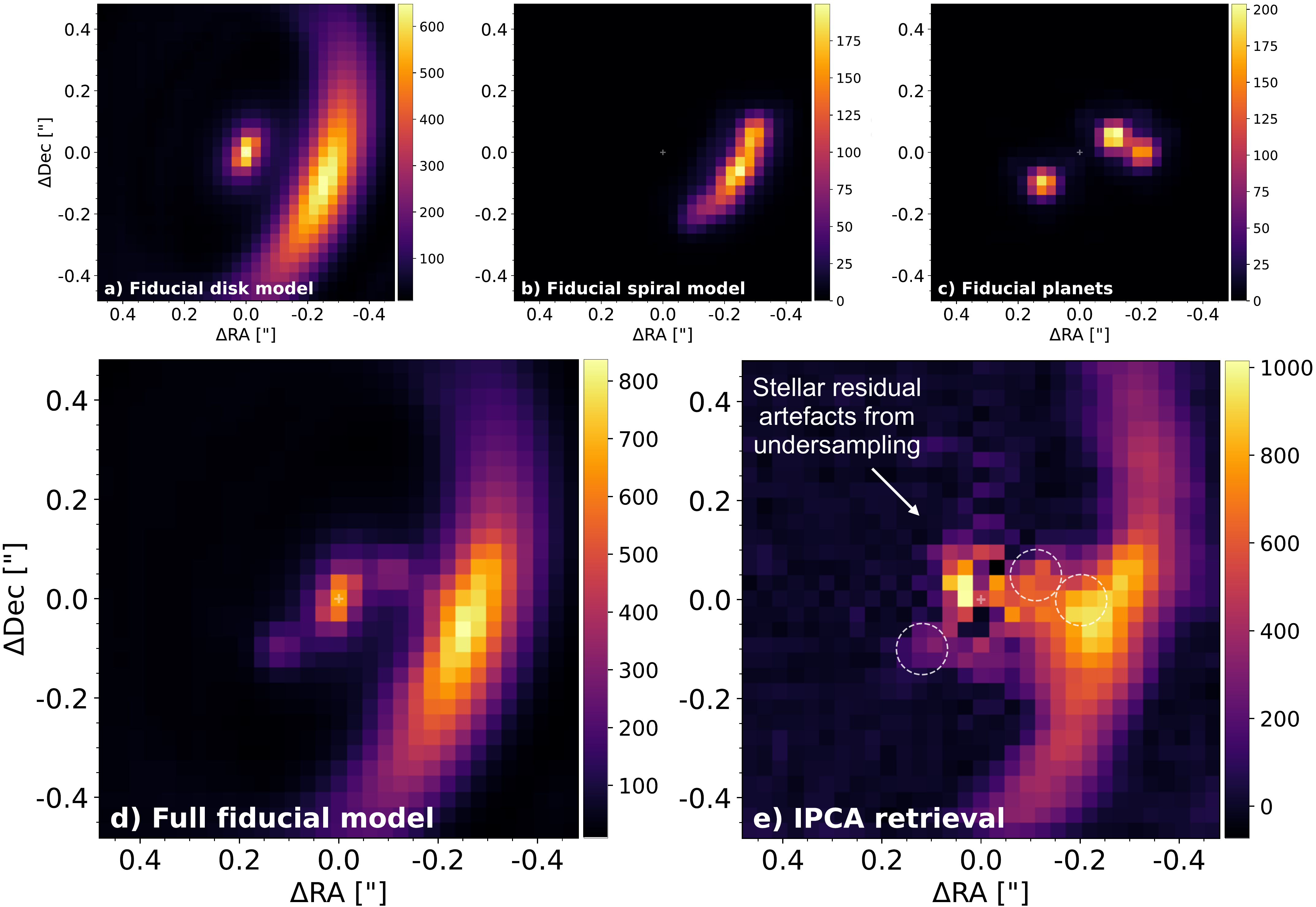}
    \caption{\new{Fiducial model of the circumstellar signals of PDS 70 ({\bf d}) compared to the image recovered by IPCA ({\bf e}) using the same reduction parameters as for the PDS 70 dataset. The different components of the model are shown in panels {\bf a-c}. This model was injected at two different roll angles separated by 5.0~\degr in the NIRCam dataset of HD~135067 at a similar contrast as the circumstellar signals of PDS 70. \vc{The protoplanet locations are indicated with dashed circles in panel {\bf e}}.}
    \vc{The recovery of all signals from the model at a similar level as injected casts confidence in the results obtained with IPCA from the PDS 70 data. 
    }}
    \label{fig:REF_test}
    \end{figure*}
    
\new{We tested the IPCA method on a fiducial dataset in order to illustrate its effectiveness at producing unbiased images of extended circumstellar signals, and to validate all our conclusions regarding the flux and morphology of circumstellar signals recovered in our IPCA images of PDS~70 in this way. For this test, we considered the only other JWST/NIRCam dataset obtained in the F187N filter with the NRCB1 detector and SUB64P subarray that was publicly available at the time of this study: an observation of reference star HD~135067 that was obtained on 2 February 2023 (Program 1902). This source does not have any known resolved circumstellar emission. We directly downloaded the calibrated \texttt{i2d} images from the MAST archive. While the number of integrations and the number of groups per integration were similar to the PDS~70 observation, the main difference was the adoption of a three-point dither pattern strategy instead of a five-point dither pattern for the observation of PDS~70. This results in a higher sensitivity to spatial undersampling effects, which are expected to be particularly prominent near the core of the PSF. As a consequence, the results of the test presented in this section should be considered as a conservative lower limit on the actual expected performance of IPCA on the PDS~70 dataset.}

\new{We designed a toy model for the circumstellar signals of PDS~70 that was composed of an outer disk, an inner disk, a spiral-like signal, and three point-source injections corresponding to protoplanets $b$ and $c$, and candidate $d$. For the outer disk, we considered the optimal model described in Appendix~\ref{sec:NEGFD}. For the inner disk, we considered similar parameters as presented in \citet{Keppler2018}, with the peak of the emission concentrated within $\sim$5~au. The two disk components are shown in Fig.~\ref{fig:REF_test}a. For the spiral-like signal, we considered a similar trace as observed in our F187N images of PDS~70 after subtraction of the optimal outer-disk model (Fig.~\ref{fig:REF_test}b). For the protoplanets, we considered similar positions and contrasts as inferred with NEGFC in Appendix~\ref{sec:NEGFC_optim} (Fig.~\ref{fig:REF_test}c). The sum of all components of the model is shown in Fig.~\ref{fig:REF_test}d. To create and inject the fiducial model into the reference cube, we relied on routines from the \texttt{fm} (forward-modeling) subpackage of VIP, in particular, its implementation of the Grenoble RAdiative TransfER tool \citep{Augereau1999, Milli2019} for the inner-disk model, its trace and fake-companion injection routines. Before injection in the reference cube data, the model was scaled to a similar contrast \vc{with respect to} the reference star as the circumstellar signals with respect to PDS 70. The signals were injected at two different angles separated by 5.0 deg in the two sets of three dithered images (i.e., the same roll-angle difference as in the PDS 70 observation).}

\new{Finally, we ran IPCA on this fiducial dataset using the same reduction parameters as were used to produce the F187N images of PDS~70. The algorithm converged within $\sim$ 400 iterations, resulting in the image shown in Fig.~\ref{fig:REF_test}e. To facilitate the comparison with the ground-truth injected model, we highlight the location of the injected protoplanets with dashed circles. Stellar residual artifacts can likely be assigned to small differences in the core of the PSF \vc{and spatial undersampling effects, which are more prominent when using a three-point instead of a five-point dither pattern. Apart from theses residuals,} we note a satisfactory recovery of both the morphology and flux level of the injected circumstellar signals. This makes us confident of the results obtained with IPCA on the PDS 70 dataset reported in this paper.
}

\section{Optimal disk model found with NEGFD}\label{sec:NEGFD}

    To be able to extract unbiased photometry for the protoplanets, the expected contribution of the disk must be removed. 
    Our goal in this work is not a full modeling of the disk, as this would involve a combined SED and disk-image fit \citep[e.g.,][]{Keppler2018,Portilla-Revelo2022, Portilla-Revelo2023}. Therefore, we decided to rely on the latest radiative transfer models of the disk produced with \texttt{MCMax3D} \citep{Min2009} and presented in \citet{Portilla-Revelo2022, Portilla-Revelo2023}, 
    as they match these combined constraints, 
    and only allow these models to vary to a small extent so that they are still compatible with ALMA and SPHERE polarized-intensity constraints. 
    Namely,
    we allowed (i) the minimum grain size $a_{\rm min}$ in the grain size distribution to vary between 0.001 and 0.05 $\upmu$m; 
    (ii) the settling parameter $\alpha$ to vary between 0.001\,$\upmu$m and 0.01\,$\upmu$m; (iii) spatial and flux scaling of the model image to vary within a small range around unity; and (iv) small linear (subpixel) and azimuthal (subdegree) shifts with respect to the center of the image, as these parameters are essentially constrained by imaging. Small variations in these parameters can lead to noticeable changes in the model images and therefore, to significant residuals after subtraction of the model. 
    We produced a grid of ten disk models in $a_{\rm min}$ (two explored values) and $\alpha$ (five explored values), and searched for the optimal model with a Nelder-Mead algorithm by interpolating model images in log-space and including free parameters for scaling and shifts. The model images were produced at the same pixel scale as the F187N and F480M images, and they were smoothed by convolution with the observed PSF. The optimal model was then found by minimizing the absolute residuals after subtraction of the model from the observed images. 
    This procedure, which we refer to as the negative fake disk technique (NEGFD), is implemented in VIP as of version 1.6.0. 

\begin{figure*}[!t]
    \centering
    \includegraphics[width=\textwidth]{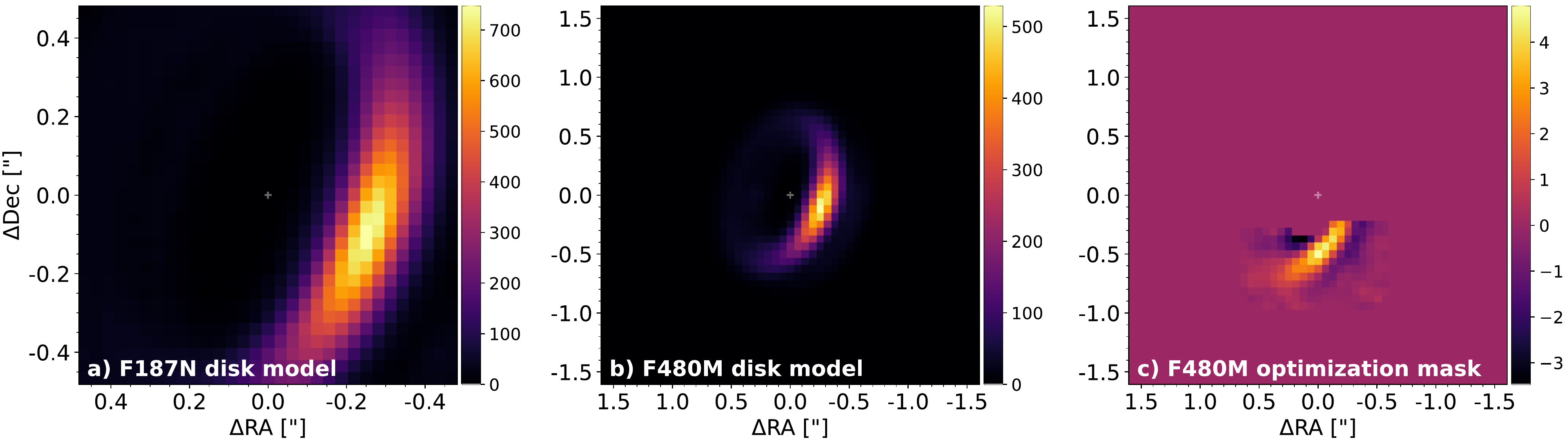}
    \caption{Best-fit radiative transfer disk models for the (a) F187N and (b) F480M images, and (c) optimization mask used to determine the optimal F480M disk model. 
    }
    \label{fig:disk_models}
    \end{figure*}

The optimal disk model was found by minimizing the sum of absolute intensity residuals in a binary mask that encompassed the location of the outer disk while excluding the location of both planets (2-FWHM aperture exclusion masks).  
We considered two potential masks. The first mask included the whole forward-scattering (i.e., brighter) side of the disk, and the second map only included the southwest part of the disk, anticipating excess signals toward the west (near planet c) and northwest part of the disk (spiral-like feature) based on previous images of the disk \citep[][]{Wang2020, Juillard2022}. As our tests using the first mask led to a mix of strong positive and negative residuals, we only consider the results obtained by minimizing residuals in the second mask, shown in Fig.~\ref{fig:disk_models}c, in the rest of this work. 
    In practice, we identified the optimal radiative transfer disk model using the F480M data because of the higher signal-to-noise ratio of the disk in these data. The optimal $a_{\rm min}$ and $\alpha$ values were found to be 0.001\,$\upmu$m and 0.01, respectively, meaning that the optimal \vc{model} is in between the radiative transfer models considered in \citet{Portilla-Revelo2022} and \citet{Portilla-Revelo2023}. 
    These parameters were then used to make a disk model prediction at 1.87\,$\upmu$m. 
    The best F187N and F480M models are shown in Fig.~\ref{fig:disk_models}a and b, respectively. 
    

Figs.~\ref{fig:NEGFD_and_NEGFC}b and e show the roll subtraction images obtained after subtracting the optimal disk model found by NEGFD from the original images (i.e., before PSF subtraction) for the F187N and F480M datasets, respectively. These are compared to the original roll subtraction images (Figs.~\ref{fig:NEGFD_and_NEGFC}a and d) using the same intensity scale. The disk subtraction performs well, in particular, for the southwest part of the disk, where the residuals reach a similar level as the residual noise level in the image. Subtraction of the optimal disk model clearly highlights the presence in the F480M image of the arm-like feature characterized in \citet{Juillard2022} in VLT/SPHERE images. This feature is indicated with a thick arrow.
We also checked how a similar NEGFD procedure performed on the IPCA images obtained with both filters by fixing the $a_{\rm min}$ and $\alpha$ to the optimal values found by NEGFC+roll subtraction. The results are shown in Fig.~\ref{fig:FinalImages}b and e, and they are discussed in Sec.~\ref{sec:spiral}.

\section{NEGFC retrieval of planet parameters} \label{sec:NEGFC_optim}

The NEGFC technique consists of finding the optimal parameters (radial separation, position angle, and contrast with respect to the star) of a directly imaged companion through the injection of fake companions with a negative flux in the input image cube (i.e., before post-processing), such that residuals are minimized in the post-processed image around the location of the planet. This forward-modeling approach is necessary to avoid the parameter estimates for the companion to be affected by self- and oversubtraction effects inherent to the post-processing algorithm used.

We used the Nelder-Mead minimization algorithm implemented in VIP and 
described in \citet{Wertz2017} to estimate the parameters for the protoplanets.
This NEGFC implementation 
offers different options in terms of figure of merit to be used in the processed image after injection of negative fake companions to identify the optimal radial separation $r_p$, position angle PA$_p$, and contrast $f_p$ of a given planet, namely minimizing (i) the sum of absolute intensity residuals in an aperture encompassing the companion, (ii) the standard deviation of intensity residuals in an aperture encompassing the companion, or (iii) the sum of absolute determinant values of the Hessian matrix calculated for each of the $n_d \times n_d$ pixels surrounding the planet location. The latter option is equivalent to minimizing the local absolute curvature, and it is more appropriate for the extraction of point-source astrometry and photometry in the presence of underlying extended signals \citep[e.g.,][]{Quanz2015}. We implemented it for this work and made it available in VIP as of version 1.6.0. 
We detail two different approaches below that involve NEGFC that we tested and compared to retrieve optimal astrometry and photometry for the protoplanets.
    
\subsection{Classical forward-modeling NEGFC}

    \begin{figure*}[!t]
    \centering
    \includegraphics[width=\textwidth]{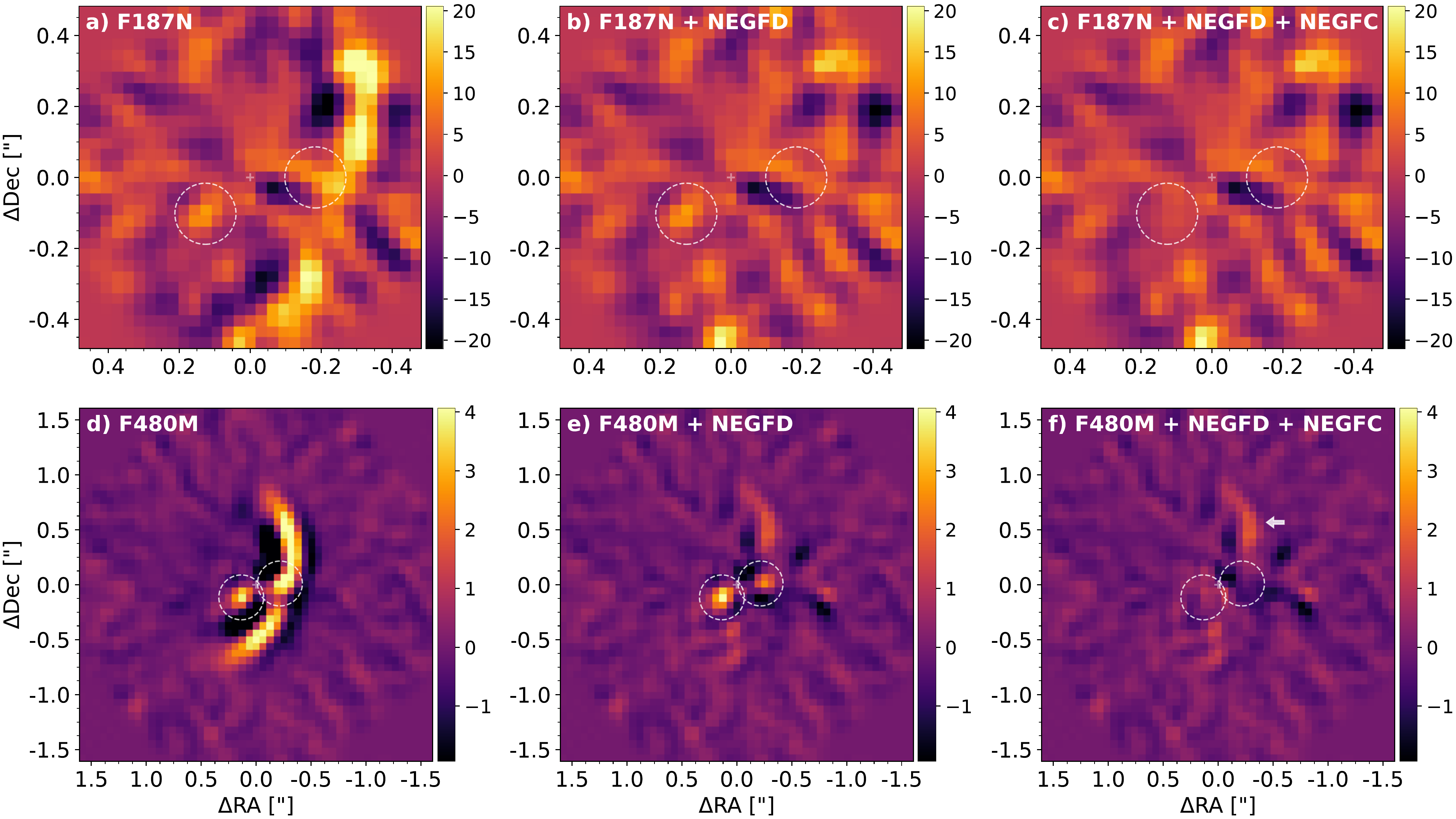}
    \caption{Images obtained at 1.87$\upmu$m (F187N; top row) and 4.80$\upmu$m (F480M; bottom row) after subtraction of the stellar PSF using roll subtraction (first column), and after additional subtraction of the best outer-disk model using the negative fake disk technique (second column; details in Appendix~\ref{sec:NEGFD}). The third column shows the 
    residuals after both disk and planet subtraction, using the parameters found with NEGFC (Appendix~\ref{sec:NEGFC_optim}). The circles indicate the predicted location of the two protoplanets based on the orbital fits presented in \citet{Wang2021}.
    }
    \label{fig:NEGFD_and_NEGFC}
    \end{figure*}

We first considered the classical forward-modeling NEGFC approach as described in \citet{Lagrange2010} or \citet{Wertz2017}, combined with roll subtraction, hereafter referred to as NEGFC+roll. We applied NEGFC+roll to the image cube where the optimal disk model found by NEGFD was removed (Appendix~\ref{sec:NEGFD}).
Regarding the choice of figure of merit, we note that different figures of merit lead to more or less reliably retrieved planet parameters depending on the S/N of the companion, the nature of the local noise, and contamination by residual extended signals (e.g., inner disk or spiral accretion stream signal). We therefore determined the optimal figure of merit on a case-by-case basis for each planet in each of our two datasets as follows. 
We used NEGFC to derive first estimates on the parameters of the companion with each of the three figures of merit, considering two subcases for the sum and standard deviation, corresponding to their calculation in either 1-FWHM or 2-FWHM size apertures, and three subcases for the Hessian-matrix determinant figure of merit, corresponding to its calculation with $n_d$ set to 1, 2, or 3. Then we injected 360 fake (positive) companions with the radial separation and contrast inferred by NEGFC, 1$\degr$ apart from each other in separate copies of the original datasets, and individually retrieved their parameters for each NEGFC subcase.
Finally, we considered the subcase that led to the smallest deviations between the retrieved and injected fake companion parameters. These deviations were also used as residual noise uncertainty on the retrieved parameters.

We did not retrieve the parameters for planet $c$ or candidate $d$ in the F187N data because they are not detected at a significant level in 
the F187N roll-subtracted image. 
For all other cases, namely planet $b$ in the F187N and F480M datasets and planet $c$ in the F480M dataset, we note that the Hessian figure of merit outperformed the other two figures of merit in terms of accuracy of the inferred astrometry. It also retrieved better the contrast of the injected companions than the sum figure of merit, which tended to overestimate their flux, likely due to residual extended signals.
For planet $b$ in the F480M data, the Hessian figure of merit is the only metric that did not significantly underestimate the radial separation of the planet compared to its expected location from the orbital fits presented in \citet{Wang2021}. The poorer performance of other metrics may be assigned to diffuse inner-disk emission (see, e.g.,~the IPCA images in Fig.~\ref{fig:FinalImages}), pushing the inferred centroid toward the center. Minimizing the determinant of the Hessian matrix is also particularly appropriate for planet c because it involves a lower sensitivity to the exactitude of the disk model inferred with NEGFD, which is subtracted from the cube before inferring the parameters of the planet. $n_d$ set to 2 or 3 led to consistent results for the Hessian figure of merit, while $n_d = 1$ was more prone to yielding outlier values.

The final uncertainties combine the residual noise uncertainties estimated with the positive fake companion injection test in quadrature with systematic uncertainties for $r_p$ and PA$_p$ and with photon-noise \vc{uncertainties} for $f_p$. We considered systematic astrometric uncertainties of 6 mas and 2 mas, for the F187N and F480M data, respectively, based on the JWST user manual. The uncertainties associated with the stellar flux are negligible in comparison to the other contributions. The results are reported in the `Roll subtraction' column of Table~\ref{tab:comparison_NEGFC}.

As a cross check, we examined the roll-subtracted images obtained after subtraction of both the disk model (NEGFD) and of the planet signals using the values retrieved by NEGFC. This is shown in the last column of Fig.~\ref{fig:NEGFD_and_NEGFC}. The residuals at the expected (circled) location of the planets are consistent with the local noise level.

\begin{table*}
\begin{center}
\caption{Comparison of the properties of the protoplanets inferred from the NIRCam F187N and F480M observations using NEGFC combined with either roll subtraction or IPCA, and the predictions made in either \citet{Wang2021} for $b$ and $c$ and \citetalias{Mesa2019a} for candidate $d$.} 
\label{tab:comparison_NEGFC}
\begin{tabular}{lcccc}
\hline
\hline
Parameter & Roll subtraction & IPCA & Final & Prediction$^{\rm (a)}$\\
\hline
\multicolumn{5}{c}{F187N} \\
\hline
\multicolumn{5}{c}{PDS~70~$b$}\\
\hline
Separation$^{\rm (b)}$ & $164.5 \pm 12.0$ & $151.9 \pm 12.1$ & $158.2 \pm 12.1$ & $155.5 \pm 1.4$ 
\\
PA$^{\rm (c)}$ & $129.0 \pm 4.1$ & $129.4 \pm 2.6$ & $129.2 \pm 4.1$ & $132.6 \pm 0.3$ 
\\
Contrast$^{\rm (d)}$ & $(3.28 \pm 0.78) \times 10^{-4}$ & $(2.50 \pm 0.56) \times 10^{-4}$ & $(2.89 \pm 0.78) \times 10^{-4}$ & $(2.4 \pm 0.1) \times 10^{-4}$ 
\\
Flux$^{\rm (e)}$ & $(8.69 \pm 2.07) \times 10^{-17}$ & $(6.62  \pm 1.48) \times 10^{-17}$ & $(7.65  \pm 2.07) \times 10^{-17}$ & $(6.3  \pm 0.3) \times 10^{-17}$ 
\\
\hline
\multicolumn{5}{c}{PDS~70~$c$}\\
\hline
Separation$^{\rm (b)}$ & -- & $203.3 \pm 23.3$ & $203.3 \pm 23.3$ & $210.1 \pm 1.0$ 
\\
PA$^{\rm (c)}$ & -- & $269.7 \pm 6.8$ & $269.7 \pm 6.8$ & $270.1 \pm 0.3$
\\  
Contrast$^{\rm (d)}$ & -- 
& $(1.37 \pm 0.42) \times 10^{-4}$ & 
$(1.37 \pm 0.42) \times 10^{-4}$
& $(4.5 \pm 0.4) \times 10^{-5}$ 
\\
Flux$^{\rm (e)}$ & -- 
& $(3.63 \pm 1.11) \times 10^{-17}$ & 
$(3.63 \pm 1.11) \times 10^{-17}$
& $(1.2 \pm 0.1) \times 10^{-17}$ 
\\
\hline
\multicolumn{5}{c}{PDS~70~d?}\\
\hline
Separation$^{\rm (b)}$ &  -- & $103.4 \pm 23.2$ & $103.4 \pm 23.2$ & $90.4 \pm 5.9$ 
\\
PA$^{\rm (c)}$ & -- & $293.0 \pm 12.7$ & $293.0 \pm 12.7$ & $284.6 \pm 1.3$
\\  
Contrast$^{\rm (d)}$ & -- & $(1.88 \pm 0.69) \times 10^{-4}$ & $(1.88 \pm 0.69) \times 10^{-4}$ & $(1.2 \pm 0.2) \times 10^{-4}$ 
\\
Flux$^{\rm (e)}$ & -- & $ (5.25 \pm 1.93) \times 10^{-17}$ &  $(5.25 \pm 1.93) \times 10^{-17}$ & $(3.4 \pm 0.6) \times 10^{-17}$ 
\\
\hline
\multicolumn{5}{c}{F480M}\\
\hline
\multicolumn{5}{c}{PDS~70~$b$}\\
\hline
Separation$^{\rm (b)}$ & $145.6 \pm 20.4$ & $141.6 \pm 10.9$ & $143.6 \pm 20.4$ & $155.5 \pm 1.4$ \\
PA$^{\rm (c)}$ & $130.4 \pm 6.7$ & $127.1 \pm 6.7$ & $128.7 \pm 6.7$ & $132.6 \pm 0.3$ \\
Contrast$^{\rm (d)}$ & $(3.93 \pm 1.50) 
\times 10^{-3}$ & $(2.96 \pm 1.02) 
\times 10^{-3}$ & $(3.44 \pm 1.50) 
\times 10^{-3}$ & $(2.8 \pm 0.6) \times 10^{-3}$ \\
Flux$^{\rm (e)}$ & $(8.13 \pm 3.11) 
 \times 10^{-17}$ & $(6.12 \pm 1.08) \times 10^{-17}$ & $(7.11 \pm 3.11) 
\times 10^{-17}$ & $(5.8 \pm 1.1) \times 10^{-17}$\\
\hline
\multicolumn{5}{c}{PDS~70~$c$}\\
\hline
Separation$^{\rm (b)}$ & $216.7 \pm 20.6$ & $219.2 \pm 12.7$ & $217.9 \pm 20.6$ & $210.1 \pm 1.0$\\
PA$^{\rm (b)}$ & $272.8 \pm 3.6$ & $276.8\pm3.7$  & $274.8\pm3.7$ & $270.1 \pm 0.3$ \\  
Contrast$^{\rm (c)}$ & $(1.55 \pm 0.30)  \times 10^{-3}$ & $(1.39 \pm 0.15) \times 10^{-3}$ & $(1.47 \pm 0.30)  \times 10^{-3}$ & $(1.0 \pm 0.3) \times 10^{-3}$\\
Flux$^{\rm (e)}$ & $(3.21 \pm 0.62) \times 10^{-17}$ & $(2.87 \pm 0.62) \times 10^{-17}$ & $(3.04 \pm 0.62) \times 10^{-17}$ & $(2.1 \pm 0.6) \times 10^{-17}$\\
\hline
\end{tabular}
\end{center}
Notes: $^{\rm (a)}$Astrometric and contrast or flux predictions based on orbital fits and best-fit atmospheric models presented in \citet{Wang2021} and \citetalias{Mesa2019a}. The contrast value reported for $d$ in \citetalias{Mesa2019a} is corrected for the expected change in total intensity-scattering efficiency based on the difference in PA with respect to the 2019 epoch. $^{\rm (b)}$Radial separation in mas. $^{\rm (c)}$Position angle measured east of north in $\degr$. $^{\rm (d)}$Contrast ratio with respect to the star and unresolved inner disk.
$^{\rm (e)}$Spectral flux density in Wm$^{-2} \upmu$m$^{-1}$. 
\end{table*}

\subsection{Direct NEGFC on the IPCA images} 

We also considered NEGFC performed directly on the IPCA images, hereafter NEGFC+IPCA, assuming that self- and oversubtraction effects are (close to) fully corrected for in these images because the IPCA images also recovered significant extended signals. It is unclear what the optimal figure of merit should be for protoplanet signals from close to or overlapping with complex extended signals, such as an inclined inner disk or a spiral accretion stream. The main source of uncertainty for the parameters of the protoplanets for NEGFC+IPCA is therefore rather associated with the correct assumption to be made in terms of the contribution from underlying extended signals at the location of the protoplanets. We therefore considered a range of figures of merit and associated parameters,
corresponding to using either the sum, standard deviation, or Hessian-matrix determinant figures of merit for a source position corresponding to either the predicted location based on \citet{Wang2021} and \citetalias{Mesa2019a} orbital fits or the visual location of a local maximum in intensity in the image. We considered two subcases corresponding to using either 1-FWHM or 2-FWHM apertures for the sum and standard deviation figures of merit, and three subcases for using either $n_d$ = 1, 2, or 3 
for the Hessian-based figure of merit.
We thus retrieved planet parameters for 14 cases in total for each protoplanet in each dataset.

These 14 cases we considered in our estimate of the final planet parameters were vetted visually upon inspection of the IPCA image after subtracting a negative planet model with parameters determined by the different figures of merit. 
We only considered visually pleasing results after subtracting the estimated protoplanet parameters from the image considering our prior knowledge of the disk (e.g., inner and outer-disk geometry), such that valid estimates typically were those from which the local intensity peak was removed while not creating a significant hole within the surrounding extended signal distribution.
The selection of different cases depended on the considered protoplanet and filter because each image and protoplanet location are affected by more or fewer surrounding circumstellar signals.
We then considered the median of the results obtained for the cases validated visually and adopted the standard deviation of these results as the uncertainty associated with the contamination by surrounding extended signals.
For the classical NEGFC approach, we added systematic astrometric uncertainties and photon noise uncertainties in quadrature to the uncertainties associated with contamination by surrounding extended signals.
Our results are reported in the `IPCA' column of Table~\ref{tab:comparison_NEGFC}.

Because of the overlap between the signals of planet $c$, candidate $d$, the spiral accretion stream, and the inner disk signals, we adopted an iterative joint-fitting strategy to estimate the parameters for planet $c$ and candidate d. We started by estimating the parameters from $c$ because it is more easily dissociated from surrounding extended signals, estimated the parameters for d in the image without the signal of $c$ (similar to Fig.~\ref{fig:FinalImages}c), then reestimated the parameters for $c$, this time, in an image from which the estimated contribution of d was removed, and so on. This procedure converged to stable estimated parameters for both sources within five iterations. We conservatively adopted for both sources the astrometric uncertainties corresponding to the largest uncertainty found among the two sources.

As a safety check, we inspected the IPCA images after subtracting the optimal parameters found for the protoplanets with the direct NEGFC procedure using the image cube from which the optimal outer-disk model was subtracted (NEGFD) to ensure that the estimates did not over- or underestimated the flux. 

    \begin{figure*}[!t]
    \centering
    \includegraphics[width=\textwidth]{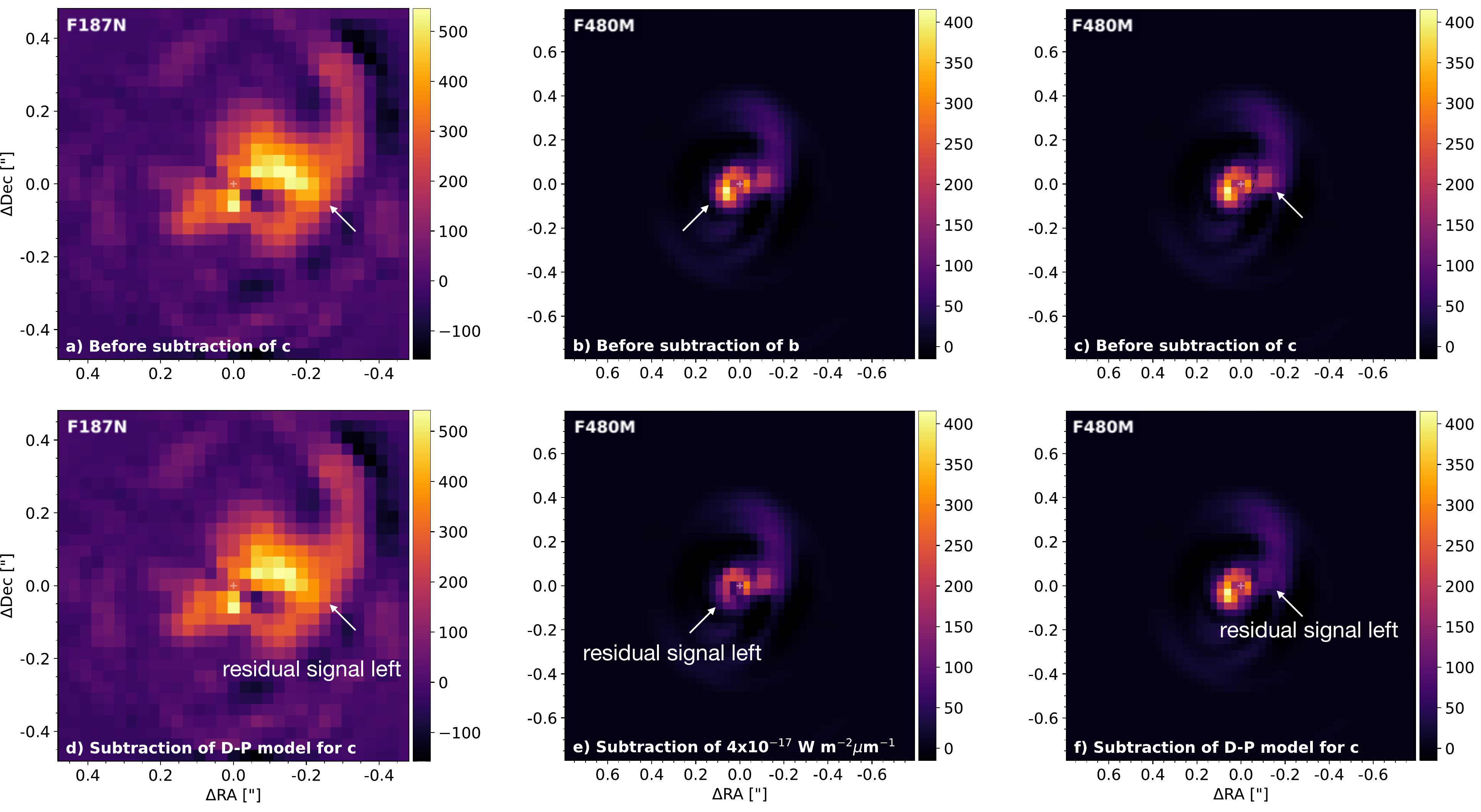}
    \caption{Images obtained at 1.87$\upmu$m (F187N; first column) and 4.80$\upmu$m (F480M; middle and right columns) before (top row) and after (bottom row) subtraction of point-source models at the location of protoplanets $c$ (first and last column) and $b$ (middle column) using fluxes \textit{different} than the optimal ones found with NEGFC. We test whether the predicted flux for planet $c$ with the DRIFT-PHOENIX model presented in \citet{Wang2021} could account for the observed flux in the F187N (panels a vs. d) and F480M (panels c vs. f) images at the location of $c$. We also show the F480M image obtained after subtracting the lower uncertainty limit from the optimal flux found by NEGFC for planet $b$ (panels b vs. e), which illustrates our uncertainty associated with the underlying inner-disk signal contribution.
    }
    \label{fig:NEGFC_safety_checks}
    \end{figure*}

\subsection{Final NEGFC results} 

Table~\ref{tab:comparison_NEGFC} shows that the astrometry and photometry extractions using NEGFC+roll and NEGFC+IPCA are consistent in all cases when retrieval of planet parameters could be made in both the roll-subtracted and IPCA images (i.e., planet $b$ with both filters, and planet $c$ in the F480M data). This suggests that IPCA recovered most of the self- and oversubtracted flux affecting roll-subtracted images. Nonetheless, neither approach is devoid of weakness for the parameter estimation. It is unclear whether IPCA did recover \textit{all} the self- or oversubtracted flux for cirumstellar signals, including the protoplanets, while on the other hand, it is unclear how much residual extended signals filtered by roll subtraction may lead to a misestimation of the NEGFC+roll planet fluxes. We therefore conservatively consider (presented in Table~\ref{tab:results}) the mean of the results obtained with NEGFC+roll and NEGFC+IPCA as our final results when both estimates are available, and we adopt the larger uncertainties of the two approaches. 



In general, the visual vetting of NEGFC+IPCA results rejected the results obtained with the sum figure of merit as it systematically led to a hole within the surrounding patch of pixels, which we interpret as likely overestimating the flux of the protoplanet alone. Only in the case of planet $b$ in the F187N data did the results appear visually satisfactory because the protoplanet is located just outward of the inner disk (although the signals from the two components appear to be connected at the resolution of our observations). In this case, it was fair to consider the possibilities that either most of the flux at the location of $b$ within our flux uncertainties is due to the protoplanet itself, or that there is nonzero contributing signal from the inner disk. These possibilities are well captured with the sum and standard deviation figures of merit, respectively. We therefore considered the median and standard deviation of the parameters retrieved for all corresponding subcases 
as our reported NEGFC+IPCA results. We also note that in this case (planet $b$ in F187N) alone, the Hessian-based figure of merit did not work properly for NEGFC+IPCA because the flux levels for $b$ and the adjacent inner-disk signals were similar.
In contrast, this figure of merit worked the best in the NEGFC+roll approach compared to the sum or standard deviation figures of merit, based on our positive fake companion injection tests, which we interpret as due to the stronger self-subtraction of (the closer-in) inner-disk signals, which causes the planet signal to stand out from it. 
Similar remarks can be made for planet $b$ in the F480M images, where the IPCA images recover inner-disk signals better and therefore tend to provide slightly closer radial separation estimates for the protoplanet due to the bias from the inner disk. 

For planet $c$ in the F187N data, the estimated parameters are affected by large uncertainties 
because many additional signals are recovered by IPCA around the planet, including the spiral accretion stream and candidate $d$, which makes the estimate of the contribution from the planet alone difficult. 
For planet $c$ in the F480M data, the NEGFC+roll and NEGFC+IPCA results are very similar. Both approaches lead to visually satisfactory results in the images obtained after subtracting the respective optimal planet parameters. 
We nevertheless note a potential shift to a larger estimated PA in the F480M data than expected from orbital fits, which may either reflect contamination by the spiral accretion stream in this lower angular resolution image or a misestimation of the orbit from earlier astrometric measurements.

\subsection{Additional tests}

To determine excess signals compared to atmospheric model predictions for the protoplanets (Sec.~\ref{sec:photometry}), we subtracted a scaled version of the observed PSF with the appropriate contrast ratio to match the flux predicted by the best-fit Drift-Phoenix model for the SED of planet $c$ presented in \citet{Wang2021} at the predicted location for the planet based on the orbital fits presented in \citet{Wang2021}. The position inferred by NEGFC in the previous sections instead of the predicted location does not change the conclusion. 
The results of this safety check are shown in the left and right columns of Fig.~\ref{fig:NEGFC_safety_checks} for PDS~70~c in the F187N and F480M filter images, respectively. The residual signals observed at the location of the protoplanet indeed suggest excess emission compared to predictions from the best-fit atmospheric models alone.

We performed a similar test for planet $b$ in the F480M data, shown in the middle column of Fig.~\ref{fig:NEGFC_safety_checks}. In this case, we subtracted the flux corresponding to the lower uncertainty of the estimate found with NEGFC (reported in Table~\ref{tab:results}), which is roughly midway between the best-fit BT-Settl model with blackbody excess and both the Drift-Phoenix and extinct EXOREM models without blackbody excess. This test suggests that our reported uncertainty is sound, and that there is thus tentative excess compared to predictions from atmospheric models without an additional blackbody contribution representative of circumplanetary disk emission. 

For comparison, the F187N and F480M images obtained after subtracting protoplanets $b$ and $c$ with the optimal parameters found by NEGFC are shown in Fig.~\ref{fig:FinalImages}c and f.


\section{Predicted and observed spirals}

    \begin{figure*}
    \centering
    \includegraphics[width=\textwidth]{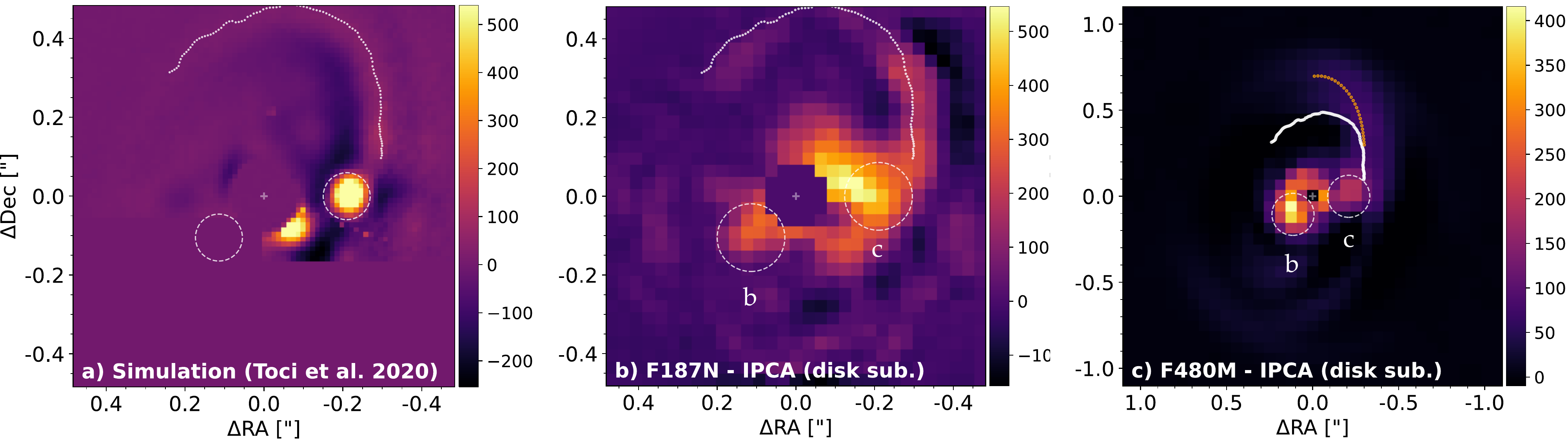}
    \caption{Comparison between the hydrodynamical simulation of the system in \citet{Toci2020}, who predicted a spiral accretion stream feeding planet $c$ (left), and our observations in the F187N (middle) and F480M (right) filters. The trace of the spiral accretion stream is identified as local radial maxima in the simulation and is shown as white dots in all images. The trace of the outer arm characterized in \citet{Juillard2022} is shown with orange dots. The part of the simulated image that is affected by subtraction artifacts is masked to show the relevant part of the image. 
    }
    \label{fig:SpiralTrace}
    \end{figure*}

Figure~\ref{fig:SpiralTrace} compares our disk-subtracted F187N and F480M images (panels b and c) to a simulated IR image predicting the spiral accretion stream associated with protoplanet $c$ (panel a). The latter is based on dedicated 3D hydrodynamical simulations made with the smoothed-particle hydrodynamics code PHANTOM \citep{Price2018}, presented in \citet{Toci2020}. The simulated image corresponds to a 
radiative transfer prediction at 2.11\,$\upmu$m of the PHANTOM simulation made with the Monte Carlo radiative transfer code MCFOST \citep{Pinte2006}, 
where a proxy for the outer-disk signal was subtracted. This proxy consists of another radiative transfer prediction for a different snapshot of the same hydrodynamical simulation, corresponding to planet $c$ being located at 180\,$\degr$ from its observed position angle. We therefore masked the bottom part of the image because it is affected by strong residuals associated with the spiral accretion stream that was subtracted from that part of the image. We measured the trace of the spiral accretion stream in the simulated image by identifying local radial maxima in the image in steps of 1\,$\degr$. The trace, shown with white dots, is then also plotted on top of the F187N and F480M images. The correspondence with the observed spiral-like signal in the F187N image and with the inner arm of the tentative fork seen in the F480M image is remarkable (better seen in Fig.~\ref{fig:FinalImages}e and f)

The arm-like signal identified and characterized in \citet{Juillard2022} is also shown with orange dots in Fig.~\ref{fig:SpiralTrace}c. This corresponds to the trace inferred from a 2021 H-band VLT/SPHERE dataset
\citep[ESO Program 60.A-9801; see more details in][]{Juillard2022}. We also note a good agreement with the outer arm 
observed in the F480M image, suggesting that it traces the same feature as observed in VLT/SPHERE images. The coarse angular resolution of the image and the different wavelength of the observation (compared to prior ground-based images in which the feature is seen) prevents a meaningful proper motion analysis of the spiral feature similar to what was done in \citet{Juillard2022}, however.

\section{Constraints on additional planets}\label{sec:contrast_curves}

\subsection{Signal-to-noise ratio maps}

    \begin{figure*}
    \centering
    \includegraphics[width=\textwidth]{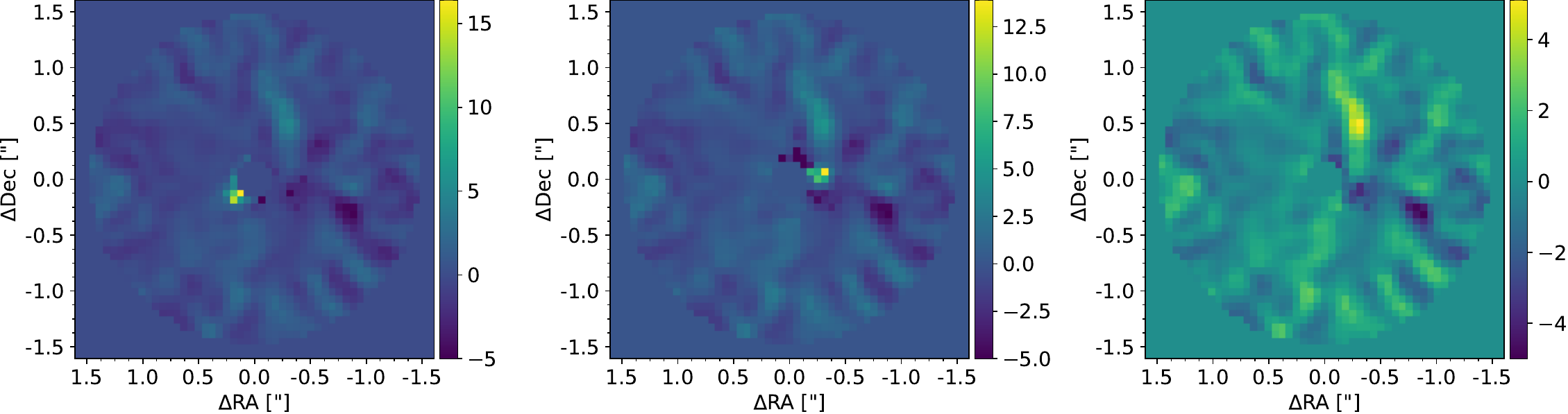}
    \caption{S/N maps of the F480M images obtained after roll subtraction and after subtraction of the disk and planets using the optimal parameters found in Appendix~\ref{sec:NEGFD} and \ref{sec:NEGFC_optim}. The left (middle) panel shows the S/N map after subtracting the disk and planet $c$ ($b$) alone, and the right panel shows the S/N map after removing the disk and both planets. Planets b and $c$ and the spiral-like feature are found at S/N values of $\sim$16.4, $\sim$13.9, and $\sim$5.1, respectively. The color bars cover S/N values ranging from -5 up to the maximum S/N value of each respective map.}
    \label{fig:SNRmaps}
    \end{figure*}
    
Figure \ref{fig:SNRmaps} shows the S/N maps obtained with roll subtraction on the image cube from which the optimal radiative disk model and protoplanets $b$ and c were all subtracted with their optimal parameters found with NEGFD (Appendix~\ref{sec:NEGFD}) and NEGFC (Appendix~\ref{sec:NEGFC_optim}), respectively.
Because the protoplanets share similar projected radii, the S/N maps are shown without planet $c$ ($b$) to evaluate the significance of planet $b$ ($c$) in the left (middle) panel. Both planets are removed in the right panel to assess the S/N of the residual signals apart from the planets. We do not detect any additional planet candidate in the outer disk. Only the spiral-like feature located in the outer wake of planet $c$ stands out at S/N$\gtrsim5$. 

\subsection{Mass sensitivity curves}

    \begin{figure*}
    \centering
    \includegraphics[width=0.495\textwidth]{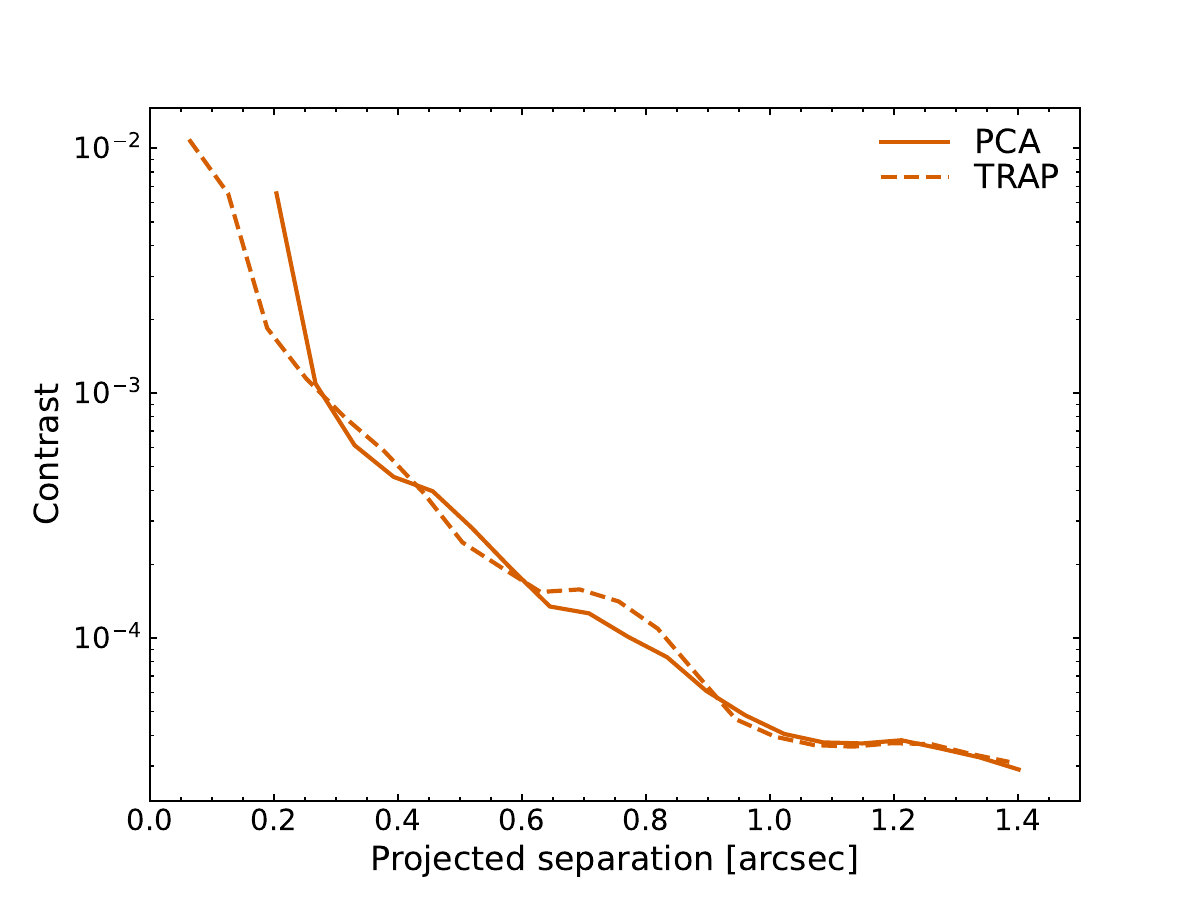}
    \includegraphics[width=0.495\textwidth]{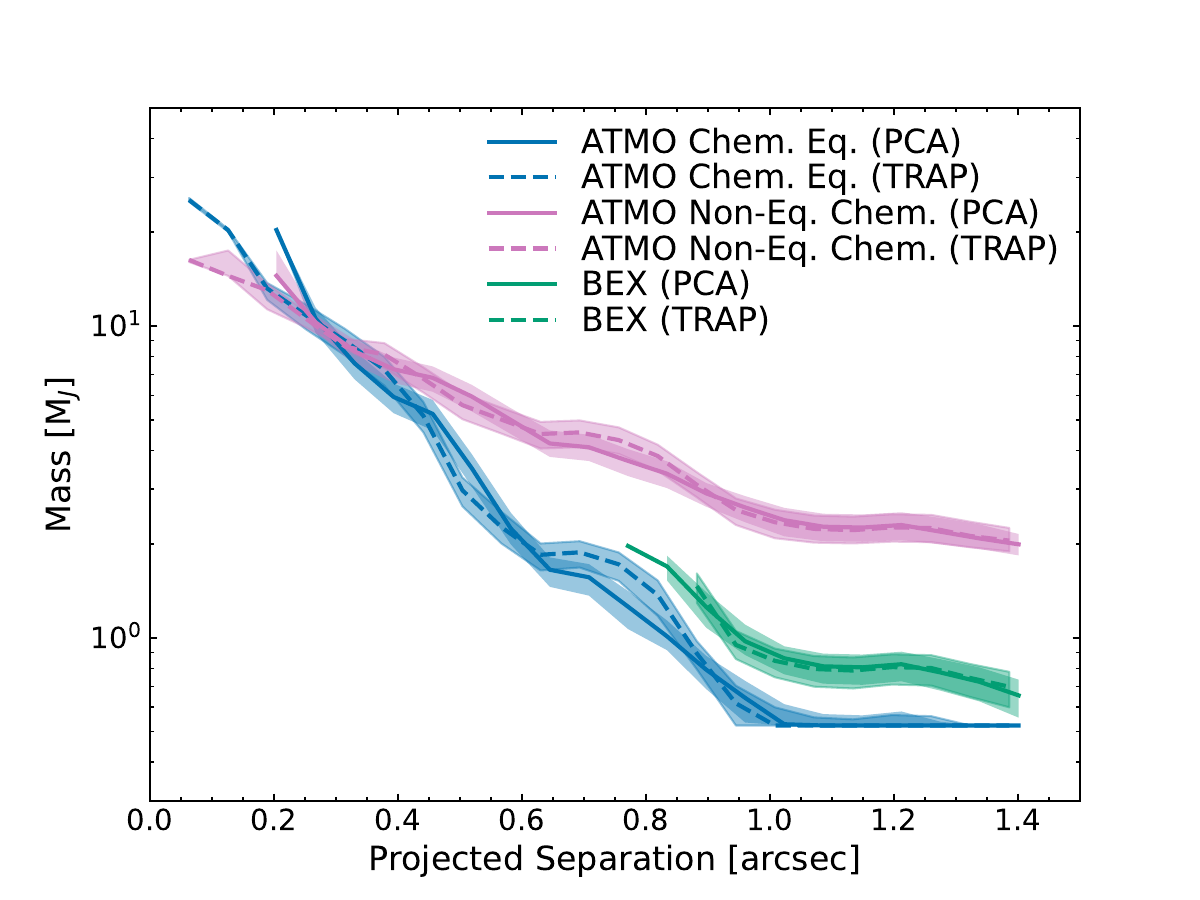}
    \caption{5 $\sigma$ contrast limits (left) and mass sensitivity (right) for the F480M observations. We show both PCA-roll and TRAP \citep{Samland2021} contrast limits and convert them into masses with a range of evolutionary models. The uncertainty on the mass sensitivity is propagated based on the uncertainty of the host star.}
    \label{fig:contrast_mass_limits}
    \end{figure*}

Fig.~\ref{fig:contrast_mass_limits} (left) shows the contrast curves achieved with PCA-roll (Appendix~\ref{sec:AltAlgos}) and TRAP \citep{Samland2021} on the image cube obtained after subtracting the optimal disk and planet models determined with NEGFD and NEGFC, respectively. A wedge was defined to encompass PA values between 10 and 160\,$\degr$ when calculating the contrast curves to avoid any significant disk residuals, including the arm-like feature, 
from biasing the contrast estimation.

We converted the achieved contrasts into mass sensitivities using the ATMO2020 models \citep{Phillips2020} with both equilibrium and nonequilibrium chemistry, and the BEX models \citep{Linder2019} assuming an age of 5.4$\pm$1\,Myr \citep{Muller2018}. These mass limits are shown in Fig.~\ref{fig:contrast_mass_limits} (right), in which the shaded region represents the 1$\sigma$ uncertainty on the mass limits due to the uncertainty in target age. The ATMO2020 evolutionary models cover masses above 0.524$M_J$, and the BEX models cover masses below 2$M_J$. There is some scatter in the calculated mass sensitivity based on the different spectra for each model assumption; models at this young age are also highly sensitive to initial conditions.
The curves show that we would have been sensitive to planets with a mass $\lesssim 2 M_J$ ($\lesssim$0.8$M_J$) at $\sim$0.6$\arcsec$ (resp.~$\gtrsim$1$\arcsec$) separation from the star, according to the BEX and ATMO (in chemical equilibrium) models; we note that we reached the bottom of the ATMO grid of models (in chemical equilibrium) in terms of planet mass. In case of nonequilibrium chemistry in the atmosphere, planets as massive as 2-4 $M_J$ may still be present, if unseen, in the outer disk.
Even higher-mass giant planets may be hidden in the outer disk \vc{if} their signal is extinct. Nonetheless, multi-Jovian mass giant planets would likely also have affected the gas density profile, which ALMA reveals to extend beyond $1\farcs5$ (i.e., beyond 170~au) in radius without any additional gaps than the gap that is carved by protoplanets $b$ and $c$ \citep{Keppler2019}.

\section[The F187N excess for planet c]{F187N excess for planet {\sf c}}\label{sec:F187N_excess}

All flux estimates shown in Figs.~\ref{fig:spec_b} and \ref{fig:spec_c} except for those from the NIRCam rely on either a model of the star plus inner disk or on absolute (spectro-)photometry of the star measured at a different epoch than the epoch of the observation that allowed the contrast measurement of the planet with respect to the star 
for the scaling of contrasts to absolute fluxes. The stellar variability plus unresolved inner-disk flux \citep{Casassus2022, Perotti2023} can thus result in significant differences between the actual photometry and the assumed photometry for the star, and may therefore accordingly affect the protoplanet flux. To illustrate the amplitude of this bias, we multiplied our measured contrasts for $b$ and $c$ by the F187N stellar flux inferred based on the 0.7-2.7$\upmu$m SpeX spectrum of PDS~70 \citep[][]{Long2018} used in other works. We show the results with light blue error bars in Figs.~\ref{fig:spec_b} and \ref{fig:spec_c}. This may thus partly account for the observed discrepancy.

Another source of misestimation for the flux of planet $c$ is the bright outer-disk edge for measurements leveraging 
angular differential imaging 
\citep{Marois2006}, and performed without prior disk model subtraction from the images.
Bright disk signals captured in the model stellar PSF 
lead to a pair of negative traces encasing the positive trace of the disk in PSF-subtracted images, similar to what can be seen in the roll-subtraction images (Fig.~\ref{fig:AltAlgos}a and e). This significantly dampens any other signal in the negative trace (e.g., planet $c$ and the accretion stream in Fig.~\ref{fig:AltAlgos}a). 
Classical NEGFC (with the sum figure of merit) or other point-source forward-modeling approaches are not tailored to properly account for self- and oversubtraction effects associated with the bright disk and may underestimate the point-source flux when it is located in a very negative disk trace like this. 
The exact amplitude of this bias is difficult to assess as it depends on the disk scattering phase function (stronger effect close to PA$_b$ where the disk is brightest), the number of principal components used for PCA-based reductions \citep[see e.g., Fig.~2 in][]{Juillard2022}, and the wavelength of the observation because the 
disk is brighter, hence the effect stronger, at shorter wavelengths. 
If the SPHERE/IFS measurements (0.9--1.6\,$\upmu$m) are underestimated, the best-fit models presented in \citet{Wang2021} may also underpredict the F187N flux, hence inflate our reported 1.87$\upmu$m excess. 

Another possibility is the inclusion of bright circumplanetary signals that are resolved in ground-based observations, but are unresolved in the lower angular resolution images of our observations, hence contributing only to our NIRCam measurements. However, the difference in angular resolution between the longest wavelengths of the SPHERE/IFS observations and our JWST/NIRCam images is minor, with an estimated FWHM about $\sim$70\% smaller for the former compared to the latter. This hypothesis therefore appears to be unlikely to account alone for the excess of a factor of $3 \pm 1$ measured in the F187N compared to atmospheric model predictions.  


Alternatively, part of the measured F187N excess may be assigned to the presence of significant Pa-$\alpha$ line emission. This hypothesis is discussed in more details in the main text (Sec.~\ref{sec:photometry}).


\section{Schematic representation of the system}

    \begin{figure}[h]
    \centering
    \includegraphics[width=0.495\textwidth]{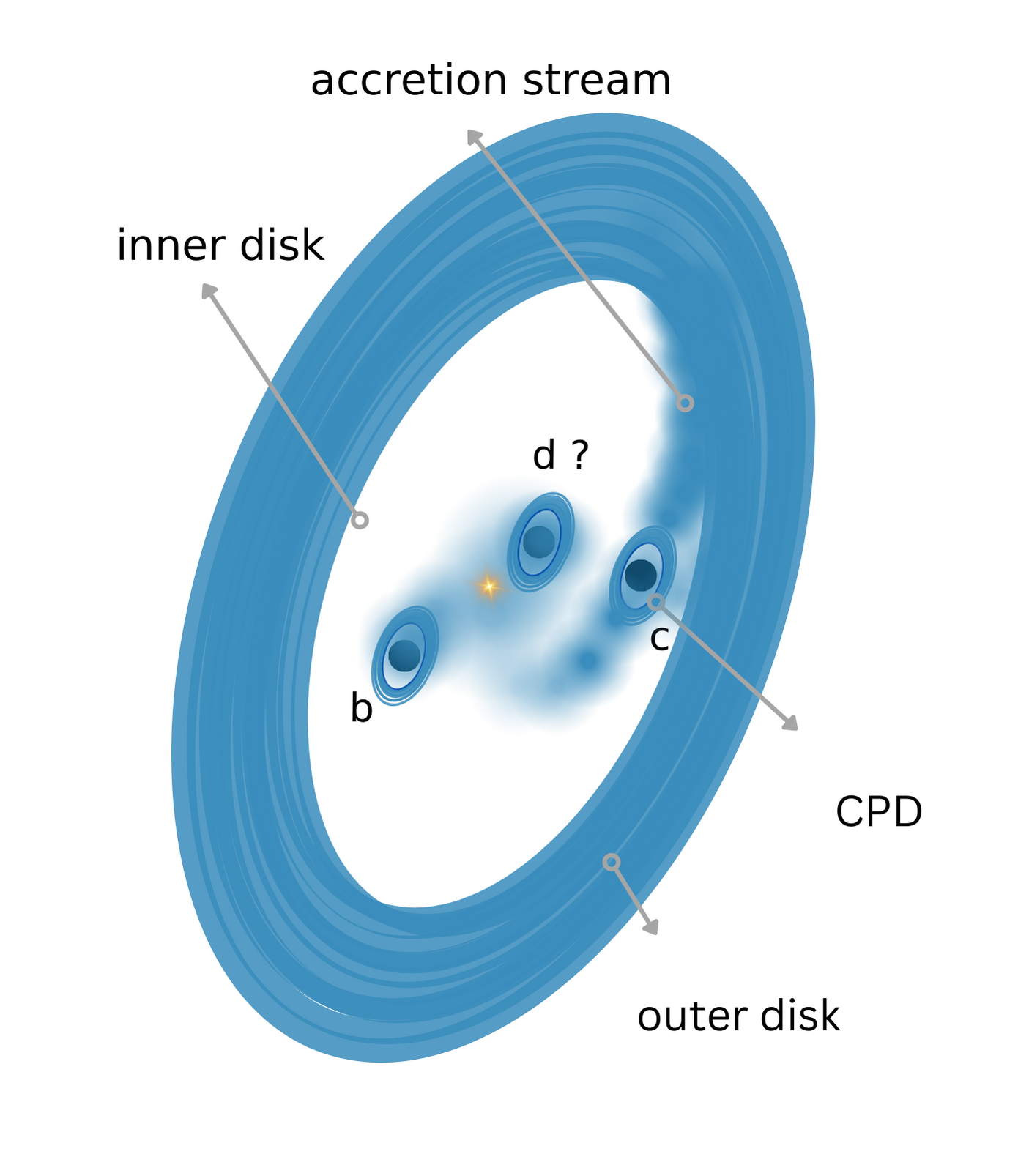}
    \caption{Schematic summary of our proposed interpretation for the main features detected in our NIRCam observation, including an accretion stream feeding the circumplanetary disk of planet $c$ and a protoplanet candidate $d$ at the edge of the inner disk.}
    \label{fig:schema}
    \end{figure}
    
In Figure~\ref{fig:schema} we summarize our knowledge about and interpretation of the different features of the PDS~70 system, drawing on \citet{Keppler2018, Keppler2019}, 
\citet{Mesa2019a}, \citet{Isella2019}, \citet{Benisty2021}, \citet{Casassus2022}, \citet{Wang2021}, and this work.

\end{appendix}
\end{document}